\documentclass[acmlarge, screen]{acmart}
 
\AtBeginDocument{%
  \providecommand\BibTeX{{%
    \normalfont B\kern-0.5em{\scshape i\kern-0.25em b}\kern-0.8em\TeX}}}

\setcopyright{cc}
\setcctype{by}
\acmJournal{IMWUT}
\acmYear{2026} \acmVolume{10} \acmNumber{3} \acmArticle{140}
\acmMonth{9} \acmDOI{10.1145/3831979}

\usepackage{subcaption} 
\usepackage{tabularx}
\usepackage{url}
\usepackage{xcolor}
\usepackage{subcaption} 
\usepackage{booktabs}
\usepackage{verbatim}
\makeatletter
\g@addto@macro{\UrlBreaks}{\UrlOrds}
\makeatother
\usepackage{multirow} %tabular cells spanning multiple rows
\usepackage{color} %colour management
\usepackage{enumitem} %better  enumerate, itemize and description

\usepackage{longtable}
\usepackage{xspace}
\usepackage{tcolorbox}
\usepackage{hyperref}

\newtcbox{\highlight}[1][magenta]{on line, arc=0pt,colback=#1!10!white,colframe=#1!50!black, before upper={\rule[-3pt]{0pt}{10pt}},boxrule=1pt, boxsep=0pt,left=6pt,right=5pt,top=4pt,bottom=3pt}

\newcommand{\notextrefrq}[1]{\hyperref[rq:#1]{\highlight{\textbf{\texttt{\textcolor{gray}{#1}}}}}}

\usepackage{todonotes}

\newcommand{\padj}{\textit{p$_{adj}$=}}

\usepackage{arydshln}

\begin{document}

\title[How Context and Individual Traits Influence Effectiveness of Different Gradual Interventions for Infinite Scrolling]{Can’t Stop: How Context and Individual Traits Influence Effectiveness of Different Gradual Interventions for Infinite Scrolling on Short-Form Video Platforms}

\author{Luca-Maxim Meinhardt}
\email{luca.meinhardt@uni-ulm.de}
\orcid{0000-0002-9524-4926}
\affiliation{%
  \institution{Ulm University}
  \city{Ulm}
  \country{Germany}
}
\affiliation{%
  \institution{Santa Clara University}
  \city{Santa Clara}
  \country{United States}
}

\author{Manuela Dragic}
\email{manuela.dragic@uni-ulm.de}
\orcid{0009-0003-8930-8502}
\affiliation{%
  \institution{Ulm University}
  \city{Ulm}
  \country{Germany}
}

\author{Mark Colley}
\email{m.colley@ucl.ac.uk}
\orcid{0000-0001-5207-5029}
\affiliation{%
  \institution{UCL Interaction Centre}
  \city{London}
  \country{United Kingdom}
}

\author{Kai Lukoff}
\email{klukoff@scu.edu}
\orcid{0000-0001-5069-6817}
\affiliation{%
  \institution{Santa Clara University}
  \city{Santa Clara}
  \country{United States}
}

\author{Enrico Rukzio}
\email{enrico.rukzio@uni-ulm.de}
\orcid{0000-0002-4213-2226}
\affiliation{%
  \institution{Ulm University}
  \city{Ulm}
  \country{Germany}
}

\renewcommand{\shortauthors}{Meinhardt et al.}

%% 170 words
\begin{abstract}
Infinite scrolling on short-form video platforms like TikTok encourages prolonged engagement and post-usage regret. Interventions aim to mitigate such behavior, but their effectiveness may depend on the interplay between intervention type, contextual factors, and individual traits. In a 7-day within-subject randomized field study (N=104), we compared a baseline pop-up and two gradually intensifying design frictions (visual and haptic). We evaluated behavioral changes and user experience using objective and subjective measures. Results showed that the pop-up was initially effective but quickly lost impact, whereas the visual gradual intervention sustained subjective ratings the longest.
Bayesian modeling revealed that self-regulation traits moderate how participants responded to the three intervention types. For participants with low impulsivity, the type of intervention had little influence on its subjective effectiveness. For participants with high impulsivity, however, differences between intervention types were substantial, with the explicit baseline pop-up being most effective compared to the novel gradual interventions. Contextual factors, in contrast, showed little influence. These findings suggest that intervention modality and individual differences in self-regulation shape intervention effectiveness.
\end{abstract}




\begin{teaserfigure}
\centering
\includegraphics[width=0.98\textwidth]{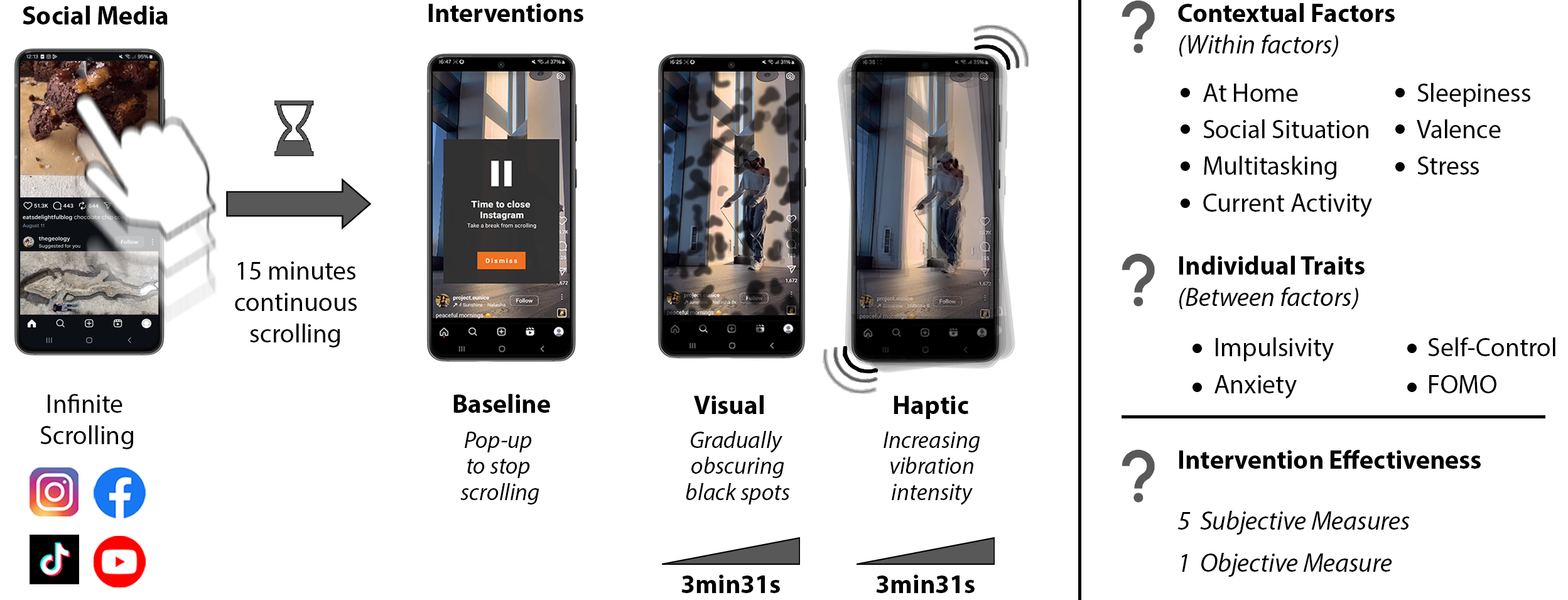}
  \caption{Overview of the 7-day field study. Participants’ continuous infinite scrolling on four major short-form video platforms (left) triggered one of three interventions (center). The baseline intervention used a pop-up, the two gradual interventions applied increasing levels of design friction via either the visual or haptic modality for up to 3\,min~31\,s. We investigated how the effectiveness of these interventions (subjective and objective) interacted with seven contextual factors and four individual traits (right).}
  \Description{The figure illustrates the study design for interventions against infinite social media scrolling. After 15 minutes of continuous scrolling on platforms such as Instagram, Facebook, TikTok, and YouTube, users encounter one of three interventions: (1) a baseline pop-up prompting them to stop scrolling, (2) a visual intervention where black spots gradually obscure the screen, or (3) a haptic intervention where vibration intensity increases while scrolling. Each intervention lasted an average of 3 minutes and 31 seconds. On the right, the figure highlights the factors examined in the study: contextual factors (e.g., location, social situation, multitasking, current activity, sleepiness, valence, stress), individual traits (e.g., impulsivity, anxiety, self-control, fear of missing out), and measures of intervention effectiveness (five subjective and one objective).}
  \label{fig:teaser}
\end{teaserfigure}
\maketitle
% Orange: #EA9924

\section{Introduction}\label{sec:intro}
Social media (SoMe) platforms like TikTok and Instagram have reshaped user engagement with digital content through mechanisms like infinite scrolling. Infinite scrolling automatically loads new content as users scroll, substantially extending screen time~\cite{Mildner.2021}. This interaction design encourages unintentional and habitual engagement~\cite{RixenIS.2023, murnane_social_2015}, often leaving users with a sense of post-usage regret~\cite{Cho.2017} and with the feeling of temporarily disconnecting from their surroundings~\cite{baughan_i_2022}. 
% This passive engagement with the content not only reduces users’ sense of control~\cite{Lukoff_2018}, but also negatively influences affective well-being~\cite{Verduyn.2015}. 
Consequently, infinite scrolling has been identified as an attention-capturing dark pattern~\cite{mongeroffarello2022towards}, designed to manipulate user behavior in ways that may conflict with their interests~\cite{Gray.2018}. Infinite scrolling on short-form video platforms such as TikTok raises particular concerns, as these platforms are associated with significantly longer screen time and more negative emotions than text-based platforms~\cite{RixenIS.2023}. Further, there is evidence that short-form video platforms decrease our ability to retain initial intentions~\cite{Chiossi.2023} and to set boundaries~\cite{viros-martin_cant_2024}. Therefore, the European Commission scrutinized TikTok's design for potential violations of the Digital Services Act, noting that it ``\textit{[...] may stimulate behavioral addictions and/or create so-called 'rabbit hole effects'}''~\cite{ECagainstTiktok}. 

Even though users are generally aware of these effects and express a desire to limit their SoMe use, they frequently struggle to follow through~\cite{ko2015nugu}. To counteract this, interventions have been designed to help people manage their SoMe use. Most approaches directly restrict access to specific apps by enforcing time limits~\cite{hiniker2016mytime} or lockout tasks~\cite{Kim.2019b}. While these strategies effectively reduce screen time, they often come at the cost of users' agency~\cite{lukoff2022designing}. Once a lockout is triggered, users are confronted with a hard boundary that they did not actively choose in that moment. According to Reactance Theory~\cite{brehm1966theory, Rains.2013}, such externally imposed restrictions threaten users’ sense of freedom, which can trigger negative emotions. Indeed, prior studies show that users frequently dismiss timers, disable reminders, or abandon lockout apps because they feel overly controlled~\cite{Okeke.2018, lukoff2022designing}. In the long term, this reactance not only undermines the intervention’s effectiveness but also reduces its acceptability, as users perceive it as intrusive or paternalistic. In contrast, \citet{lu_interactout_2024} showed that making common gestures like taps or swipes slightly more difficult to perform can more subtly nudge users to reduce smartphone use. This principle of design friction~\cite{cox_design_2016} introduces micro-boundaries ``\textit{[...] that provide a small obstacle prior to an interaction that prevents us rushing from one context to another}''~\cite[p. 1392]{cox_design_2016}. These interruptions, though minor, create moments for reflection, prompting users to reconsider their behavior~\cite{benford_uncomfortable_2012, cox_design_2016, mejtoft_design_2019}. Hence, design friction has been successfully applied to mitigate SoMe overuse~\cite{haliburton_longitudinal_2024} and, more specifically, scrolling behavior~\cite{ruiz_design_2024}. One challenge with design friction is finding the right level of friction~\cite{kim2019goalkeeper, ulriklyngs2022goldilocks}. Too little friction is easily ignored; too much can provoke reactance and abandonment of the intervention~\cite{lukoff2022designing}. Gradually increasing intensity offers one potential, novel design strategy to find the best level of friction over time, as it allows interventions to become harder to ignore the longer scrolling continues. This motivates our first research question (RQ1):

\begin{quote}
      \textbf{RQ1: How do users experience interventions over time as friction intensity increases?}
\end{quote}

\noindent Whether such strategies succeed depends not only on how friction is designed, but also on where and when it is encountered~\cite{purohit_2019}, as their effectiveness is likely to vary across contexts. For instance, during work hours, subtle interventions may suffice as social norms already discourage phone use~\cite{Miller-Ott.2017, FORGAYS2014314}, whereas during leisure time, stronger interventions may be necessary~\cite{RixenIS.2023}. \citet{meinhardt_scrolling_2025} also argued to systematically explore ``\textit{various interventions to determine the most effective ones for specific contexts}''~\cite[p. 12]{meinhardt_scrolling_2025}. Beyond context, prior work found that individual traits such as users' ability for self-regulation predict smartphone addiction~\cite{gokcearslan_modelling_2016} and problematic SoMe usage~\cite{li_mediating_2025}. In particular, low self-control, high fear-of-missing-out (FOMO), and high impulsivity can lead to problematic SoMe use~\cite{simsir-gokalp_self-control_2024, koc_relationships_2023, Guo2022Applying}. While the relationship between these traits and problematic use is well-known, the moderation between them and the effectiveness of specific interventions remains largely unexplored. \citet{mark_effects_2018} provide early evidence that this relationship matters in the workplace, where distraction interventions helped people with low self-control focus but caused stress for those with high self-control. Whether individual traits and contextual factors similarly moderate the effectiveness of different types of interventions for infinite scrolling during short-form video platforms is the central question of this work. This motivates our RQ2:

\begin{quote}
      \textbf{RQ2: How do \textit{contextual factors} and \textit{individual traits} moderate users' behavior change and experience of different interventions during infinite scrolling?}
\end{quote}

\noindent To answer both RQs, we operationalized intervention effectiveness along two dimensions: objective and subjective. Objective effectiveness was measured using \textit{responsiveness}~\cite{meinhardt_scrolling_2025}, defined as the time to stop infinite scrolling after an intervention occurred. However, prior research shows that reducing screen time alone can provoke negative reactions and relapse into old habits~\cite{Okeke.2018}. 
% Further, scholars criticized screen time as an insufficient success metric for interventions~\cite{Lukoff_online, hiniker2016mytime, Almoallim_2022}. 
Therefore, we also considered subjective effectiveness as the weighted combination of \textit{reactance}~\cite{Ehrenbrink.2020}, \textit{goal alignment}~\cite{kai_internal}, \textit{usefulness}~\cite{Venkatesh2000}, \textit{agency}~\cite{kai_internal}, and \textit{satisfaction}~\cite{kai_internal}. In our 7-day field user study ($N=104$), participants installed an Android application that monitored infinite scrolling behavior on TikTok, Instagram, Facebook, and YouTube Shorts. Once a scrolling session exceeded 15~min (in line with~\cite{meinhardt_scrolling_2025, RixenIS.2023, Terzimehic.2022b}), the app randomly triggered one of three interventions: The baseline intervention showed a pop-up that encouraged users to take a break from scrolling. The other two interventions gradually intensified either visual or haptic friction over 3~min and 31~s~\cite{meinhardt_scrolling_2025} (see \autoref{fig:teaser}). In the visual condition, the content became progressively more challenging to view, while in the haptic condition, the phone increasingly vibrated.
After users stopped scrolling due to the intervention, they rated their subjective intervention's effectiveness and reported their current context (e.g., location, valence, stress). Before the study, we collected the participants' traits for \textit{impulsivity}, \textit{FOMO}, \textit{anxiety}, and \textit{self-control}.

Our results show that the baseline pop-up intervention yielded the highest objective effectiveness, but its subjective experience diminished rapidly if users did not act immediately, suggesting that simple reminders work primarily for those already motivated to disengage. In contrast, the visual gradual intervention maintained stable subjective ratings throughout its increase in intensity, potentially supporting longer-term effectiveness by avoiding abrupt drops in the intervention’s salience. 
% The haptic gradual intervention fell in between, initially effective but declining sooner, likely because increasing vibration intensity became intrusive. 
We found that individual traits related to self-regulation, particularly self-control and impulsivity, moderate the relationship between our intervention types and both subjective and objective effectiveness. By contrast, contextual factors showed only limited interaction effects. For participants with high self-control or low impulsivity, the intervention type mattered less, likely because they could disengage from infinite scrolling without much external support. In contrast, participants with low self-control or high impulsivity responded most effectively to the baseline intervention. Its explicit prompt to \textit{take a break} may provide the clear external signal these users need, whereas the gradual interventions might be too subtle for those with high impulsivity. 
While we tested only three different intervention types, our results suggest that future work should move beyond one-size-fits-all interventions. We suggest that interventions should be carefully designed to tailor them to individual differences. Further, intervention designers should be careful about what they optimize: objective behavior change vs. subjective experience. Our results reveal a trade-off between these two measures. Interventions that cut scrolling fastest can feel frustrating, whereas more acceptable designs may delay disengagement. Effective interventions should balance these goals, or adapt which goal they prioritize.

\vspace{1em}

\noindent\fcolorbox{orange}{orange!30}{\textbf{Contribution Statement}~\cite{Wobbrock.2016}}

\vspace{0.8em}

\noindent\textbf{Empirical study that tells us about people $\mid$ Artifact.} We contribute a 7-day field study ($N=104$) comparing a baseline pop-up with two novel gradually intensifying interventions (visual vs. haptic) to mitigate infinite scrolling. We demonstrate that, within the scope of our study, intervention effectiveness depends not only on our tested designs but also on individual differences in self-regulation ability, such as impulsivity and self-control. 
% This suggests that digital well-being interventions might not be designed as one-size-fits-all. Instead, they should balance objective behavior change with subjective user experience and adapt to individual differences in self-regulation.

\section{Related Work}
This section reviews interventions aimed at reducing SoMe overuse and promoting digital well-being by limiting smartphone use. We also examine prior research on how contextual factors and individual traits influence problematic phone usage and behavior change. 
% Problematic smartphone use has been extensively studied~\cite{Thomee.2018, bashir2017effects, Panova.2018, Lee.2014, Sha.2021, Verduyn.2015}, typically from two perspectives: one views excessive phone use as exhibiting addictive behavior~\cite{lin2015time}; the other considers whether certain interface designs are inherently problematic~\cite{Mildner.2021, Mildner.2023}. With this regard, 
\citet{Chen.2023} classified moments of killing time and found that especially sessions of prolonged infinite scrolling were likely labeled as time-killing. In response, researchers have proposed interventions to reduce SoMe and smartphone use, which we briefly summarize below.

\subsection{Interventions for Limiting Social Media Use}
Digital interventions to limit SoMe use can be categorized into internal and external~\cite{Purohit.2023, kai_internal}. Internal interventions are embedded in the app, such as removing the newsfeed to reduce exposure to endless content~\cite{Purohit.2023}. External interventions operate at the device level, modifying smartphone behavior without altering app functionalities. These external interventions vary in intrusiveness and can be grouped into four main types~\cite{mongeroffarello2019race}: First, usage timers help users reflect on and potentially adjust their habits~\cite{hiniker2016mytime}. Second, persuasive interventions use prompts or notifications to encourage mindful engagement~\cite{Purohit2021,2023cocreation}. Third, break reminders aim to interrupt usage and direct users toward more meaningful activities, such as breathing exercises~\cite{Terzimehic.2022b, haliburton_longitudinal_2024}. Lastly, phone blockers restrict access entirely or increase usage difficulty, providing support for individuals struggling with self-regulation~\cite{Kim.2019b}. As an alternative to such interventions, timeboxing structures phone use into bounded sessions with self-set time allowances rather than restricting access outright~\cite{park2021goldentime}.

While internal and external interventions reduced SoMe use, internal interventions are more effective on passive SoMe features such as infinite scrolling~\cite{orzikulova_finerme_2023}. \citet{lee_purpose_2025}, who tested an intervention that disabled attention-capturing dark patterns such as infinite scrolling, showed 21\% less distraction when these dark patterns were disabled.
Most interventions employ a rather intrusive approach, delivering them as a sudden interruption of the user's interaction. In contrast, the design friction approach advocates for small obstacles prior to the interaction to create moments of reflection to change behavior~\cite{benford_uncomfortable_2012, cox_design_2016, mejtoft_design_2019}. Recent work proposed applying design friction specifically to scrolling behavior: \citet{ruiz_design_2024} required users to interact with each post before being allowed to scroll further. Participants reported that this increased their attention to individual posts. However, they also expressed frustration due to the repetitive interaction required. Despite their variety, many intervention mechanisms rely on uniform application across users, overlooking contextual and individual factors that could influence the interventions' effectiveness. %The following section explores how contextual factors shape digital media use and prepares the ground for designing more context-aware interventions.

\subsection{The Role of Context in Digital Media Use}
Contextual factors, such as environmental setting, mobility, social interactions, multitasking, and distractions, can influence digital media use~\cite{Akpinar.2023}. For instance, \citet{bohmer_falling_2011} found that users consume news content in the morning and shift to SoMe content in the evening. Further, \citet{xu_identifying_2011} observed higher mobility among users of social networking and gaming apps.
A large-scale behavioral study confirms these trends~\cite{Hintze.2017}. They further showed that smartphone use is nearly twice as high at home compared to the office~\cite{Hintze.2017}. \citet{RixenIS.2023} also observed an 18\% drop in infinite scrolling during work-related activities compared to leisure. They also found that infinite scrolling is often used as a coping mechanism for negative emotions or procrastination. Beyond temporal and locational influences, social context plays a crucial role. \citet{Do.2011} found that the presence of nearby Bluetooth devices was associated with increased smartphone use. Conversely, \citet{Weber.2020} showed that people were less likely to use their phones during group dining.
These findings highlight the dynamic nature of digital media usage and introduce the importance of context for changing the behavior of digital media consumption~\cite{thomas2021systematic, karppinen2018opportunities}. Acknowledging this, \citet{Ding.2016} emphasized the crucial role of time and location when it comes to setting reminders to change behavior. They argue that ``\textit{[...] context information plays a very important role in increasing the effectiveness and reducing the annoyingness of reminders [interventions]}''~\cite[p. 7]{Ding.2016}. 

\subsection{Context-Aware and Tailored Interventions}
Recent work argues for moving beyond one-size-fits-all strategies by tailoring SoMe interventions to users’ situational context~\cite{alberto_2021, alberto_2023, RixenIS.2023, Purohit.2019, Okeke.2018framework, Sobolev.2021, VandenAbeele.2021b}. For example, \citet{orzikulova_time2stop_2024} designed adaptive just-in-time interventions based on phone usage, physical activity, time of day, location, and social setting. By using machine learning to predict optimal intervention moments, their system improved intervention accuracy by 33\% compared to non-adaptive interventions. Similarly, \citet{wu_mindshift_2024} showed that mental states (boredom, stress, or inertia) shape how users respond to interventions. They generated personalized pop-ups using large language models
% based on strategies such as comforting or scaffolding, 
and found that integrating mental state increased acceptance by up to 22.5\%. 
% Beyond SoMe, adaptive approaches have also proven effective in other behavioral domains. For instance, \citet{korinek_adaptive_2018} developed a walking intervention that dynamically adjusted daily step goals and rewards based on participants’ prior performance. This adaptive design significantly improved adherence compared to static goal-setting, demonstrating the potential of tailoring intervention intensity to individual progress over time.
\citet{meinhardt_scrolling_2025} investigated how contextual factors influence user responses during infinite scrolling. They found that low valence, combined with being at home, led to users disengaging from scrolling more quickly. In contrast, when sleepy, they were more likely to accept interventions but were less likely to act on them. However, their study focused on a single intervention type and did not examine whether individual differences also moderate intervention effectiveness. This limitation is important because individual traits are known predictors of problematic SoMe use~\cite{Guo2022Applying, simsir-gokalp_self-control_2024, koc_relationships_2023}. Moreover, interventions themselves are not uniformly effective. Distraction blockers, for example, helped people with low self-control to focus but caused stress for those with high self-control~\cite{mark_effects_2018}. If such traits shape both usage patterns and reactions to restrictions, they will also likely determine how individuals perceive and respond to interventions during infinite scrolling.

%Building on these insights from prior research, 
Our work addresses this gap by systematically comparing how three intervention types differ in objective behavior change and subjective experience, and by testing whether individual traits and contextual factors moderate these differences. Unlike prior work that established relationships between individual self-regulation traits and problematic phone use~\cite{gokcearslan_modelling_2016}, our study directly examines whether these traits also determine which intervention type is most effective for a given user. 
% Hence, we move towards a more adaptive approach to digital well-being interventions that selects the most appropriate intervention based not only on the user's current context but also on their stable individual traits.

\section{Contextual Factors and Individual Traits for the User Study} \label{sec:peronality_contextual_factors}
To account for the diverse range of influences on intervention effectiveness during infinite scrolling, we identified eleven factors grounded in prior research. These include both between-subject individual differences and within-subject contextual factors. %These factors were investigated in our user study and are described below.

\subsection{Contextual Factors \textit{(Within Factors)}}
We included seven contextual factors (within-factors) in our study: \textit{current activity}, \textit{social situation}, \textit{location (at home)}, \textit{multitasking}, \textit{valence}, and \textit{sleepiness}. These factors influence how users engage with their devices and respond to interventions during infinite scrolling~\cite{Purohit.2019, RixenIS.2023, meinhardt_scrolling_2025}.
For example, interventions may be more effective when users are engaged in activities like work rather than during leisure~\cite{RixenIS.2023}, or when social norms discourage phone use, such as in social gatherings~\cite{Miller-Ott.2017, Misra.2016}. Similarly, usage \textit{at home} tends to be longer~\cite{Hintze.2017}; therefore, during this context, the intervention effectiveness may be reduced~\cite{meinhardt_scrolling_2025}. Further, \textit{multitasking} (e.g., eating or watching TV while being on the phone) creates cognitive distractions to redirect attention away from the phone, especially when being \textit{at home} or during low \textit{valence}~\cite{meinhardt_scrolling_2025}. Regarding the latter, \citet{RixenIS.2023} found that extensive scrolling sessions were linked to lower \textit{valence}, suggesting a negative emotional impact. Since users often cope with negative emotions via smartphones~\cite{Diefenbach.2019}, emotional state may also influence the intervention's effectiveness.

We considered including time of day as a contextual factor but ultimately excluded it, in line with \citet{meinhardt_scrolling_2025}. Due to its periodic nature and strong individual variability in daily routines (e.g., between shift workers and students), time of day is a poor proxy for user state in our modeling. Instead, we included \textit{sleepiness} as a more individual temporal indicator. Sleepiness varies throughout the day, can be modeled linearly, and meaningfully predicts user behavior~\citet{meinhardt_scrolling_2025}. 
% Prior research also links excessive smartphone use with poor sleep quality~\cite{YANG2020112686, Christensen.2016} and increased daytime sleepiness~\cite{Nathan.2013}, which lowers reactance towards interventions~\cite{meinhardt_scrolling_2025}.
In addition to the above-mentioned six contextual factors, we included \textit{stress}. As \citet{VandenAbeele.2021b} mention \textit{stress} as a key determinant of digital well-being, it is reasonable to assume that \textit{stress} may also shape how users perceive and respond to interventions.

\subsection{Individual Traits \textit{(Between Factors)}}

Digital well-being is also shaped by individual differences, particularly traits such as \textit{impulsivity}, \textit{self-control}, \textit{anxiety}, and \textit{FOMO}~\cite{VandenAbeele.2021b}, all of which have established links to problematic SoMe use. \textit{Impulsivity} is the tendency to act in favor of immediate rewards without deliberation~\cite{nigg_annual_2017} and is strongly linked to SoMe addiction~\cite{Guo2022Applying}. \textit{Self-control} describes the capacity to override such urges in favor of long-term goals~\cite{tangney_high_2004}, with low self-control relating to more problematic use~\cite{simsir-gokalp_self-control_2024}. Both constructs are typically strongly negatively correlated~\cite{mao_self-control_2018}, yet they are not redundant. Impulsivity ``\textit{is not merely a lack of self-control, but rather a manifestation of relatively high levels of appetitive motivation for which self-control may serve as a buffer [...]}''~\cite[p.~74]{mao_self-control_2018}.
\textit{Anxiety} reflects a disposition toward anticipatory apprehension, or worry, about perceived future threats~\cite{craske_anxiety_2016} and relates positively to problematic SoMe usage~\cite{shannon_problematic_2022}. Lastly, \textit{FOMO} describes the apprehension of missing rewarding experiences others might be having~\cite{przybylski_motivational_2013} and it increases the risk of problematic phone use~\cite{koc_relationships_2023}. 

While the relationship between these four traits and problematic phone and SoMe usage is well established, less is known about whether they also moderate the effectiveness of interventions designed to mitigate such behaviors. Initial evidence from \citet{mark_effects_2018} suggests that individual differences in self-control shape how users respond to an intervention, but this question has not been examined across different intervention types for infinite scrolling on SoMe. Based on their influence on SoMe usage, we included \textit{impulsivity}, \textit{anxiety}, \textit{self-control}, and \textit{FOMO} as between-subject factors to examine their moderating role in intervention effectiveness during infinite scrolling.

\section{User Study}
To examine how contextual factors and individual differences moderate objective and subjective effectiveness of three intervention types during infinite scrolling, we conducted a 7-day within-subject field study with $N=104$ participants. This duration was chosen to capture a sufficient range of intervention encounters across varied situational contexts rather than to demonstrate lasting behavior change, which would require longer observation periods~\cite{haliburton_longitudinal_2024}. A week also spans a full cycle of participants' weekday and weekend routines, so that recurring daily situations are represented at least once in the data. We focused on four of the most widely used short-form video platforms in the United States in 2023 (Instagram, TikTok, Facebook, and YouTube Shorts)~\cite{StatistaSNS}, aligned with prior work~\cite{RixenIS.2023, meinhardt_scrolling_2025}.

\subsection{Apparatus}\label{sec:apparatus}
Our study was built on prior work, which open-sourced their application to track infinite scrolling~\cite{meinhardt_scrolling_2025}. We extended this work by adding two novel, gradually intensifying interventions (see \autoref{sec:interventions}). The tracking mechanism remained the same: using Android’s Accessibility Service~\cite{GooglePlayStore.AccesabilityService}, the app accessed the content variable of each app’s hierarchy to identify infinite scrolling. Active engagement with SoMe, such as messaging or content creation, was excluded. For example, if a user switched from Instagram’s “Reels” tab to the messaging interface, the content variable shifted, which was interpreted as a scrolling interruption. This focus on passive scrolling aligns with findings that interventions are more effective when targeted at specific in-app features rather than overall SoMe time~\cite{orzikulova_finerme_2023}.

A scrolling session was defined as uninterrupted feed consumption until the user either switched to a non-scrolling activity or closed the app. After 15\,min of continuous scrolling, the app randomly triggered one of the three interventions with equal probability (1/3 each). Randomization occurred independently at each intervention moment, so participants could receive the same or different interventions across sessions (see \autoref{sec:interventions}). We used the 15\,min threshold as done in previous works~\cite{meinhardt_scrolling_2025} based on the findings that negative emotions from smartphone use often emerge after 10-20\,min~\cite{Terzimehic.2022b} and that sessions exceeding 10\,min are typically dominated by infinite scrolling~\cite{RixenIS.2023}. 
When users stopped scrolling after an intervention, a short questionnaire was presented to collect contextual information (\autoref{sec:measure_context_factors}) and assess users' subjective intervention effectiveness (\autoref{sec:intervention_effectiveness}). This event-based Experience Sampling Method (ESM) was chosen over fixed-interval ESM because it captures context immediately after disengagement from infinite scrolling, improving ecological validity and response rates~\cite{vanBerkel.2019}. However, this method only captures the participant's context at the moment they decide to stop scrolling. Hence, we missed data from those who continued scrolling. Despite this, we chose to collect subjective measures after infinite scrolling ended rather than during scrolling because interrupting the activity would have disrupted the scrolling behavior we aimed to observe. This approach is consistent with other ESM-based SoMe studies (e.g., \cite{Cho.2021, RixenIS.2023, Chang.2015, Bayer.2018}). 

We also recorded the time that passed between the intervention occurring and the user stopping their scrolling behavior. We used his duration (\textit{responsiveness}) to calculate the users' objective effectiveness of the intervention (see \autoref{sec:intervention_effectiveness}).

\subsubsection{Interventions}\label{sec:interventions}
\begin{figure*}[ht!]
\centering
\small
    \begin{subfigure}[c]{0.2\linewidth}
        \includegraphics[height=3.8cm]{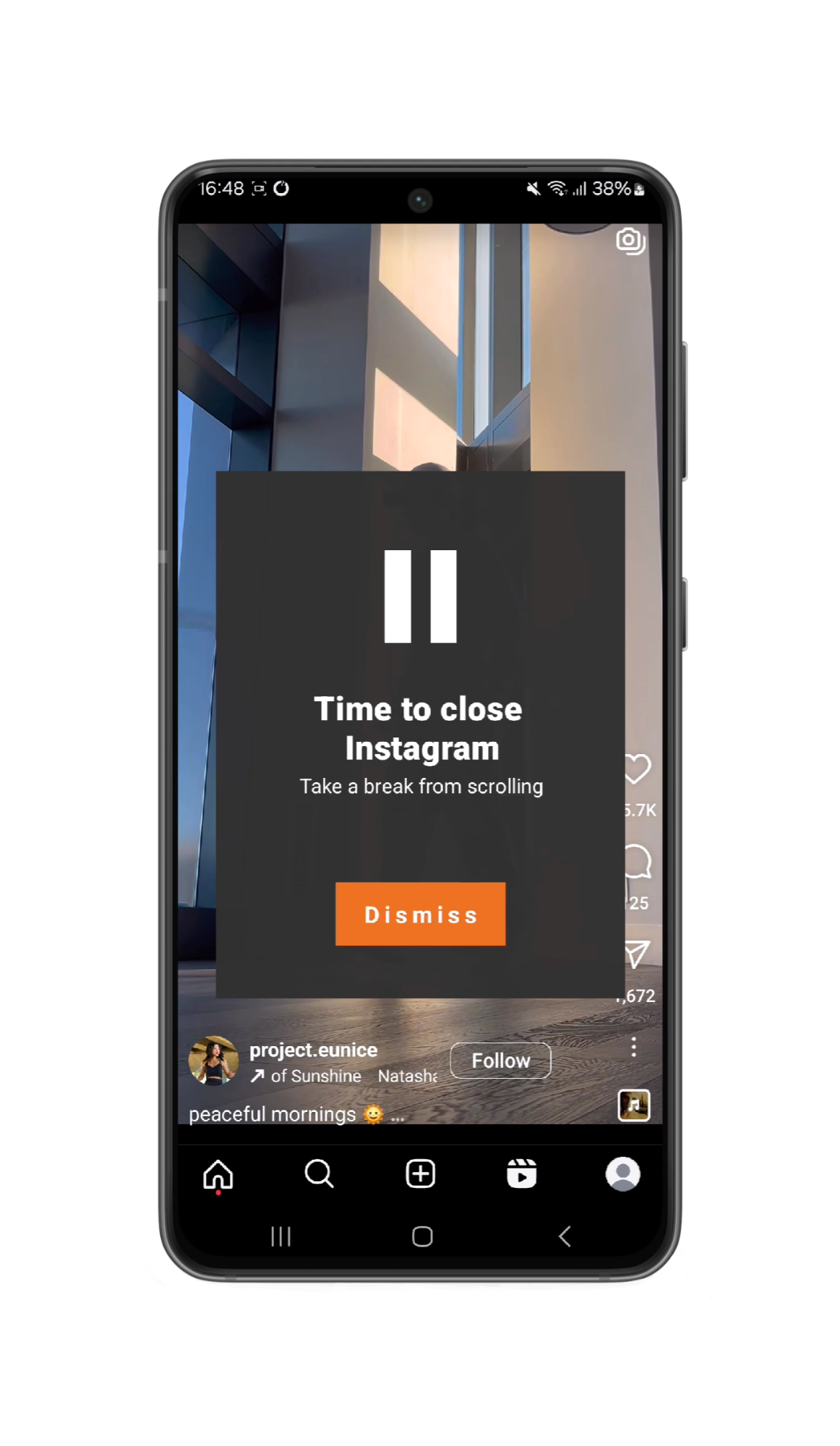}
        \caption{Baseline Intervention}\label{fig:baseline_intervention}
        \Description{Baseline Intervention: A smartphone screen displays a pop-up message reading “Time to close Instagram, Take a break from scrolling” with a pause icon above and a button labeled “Dismiss”.}
    \end{subfigure}
    % \hspace{1.4cm}
    \begin{subfigure}[c]{0.38\linewidth}
        \includegraphics[height=3.8cm]{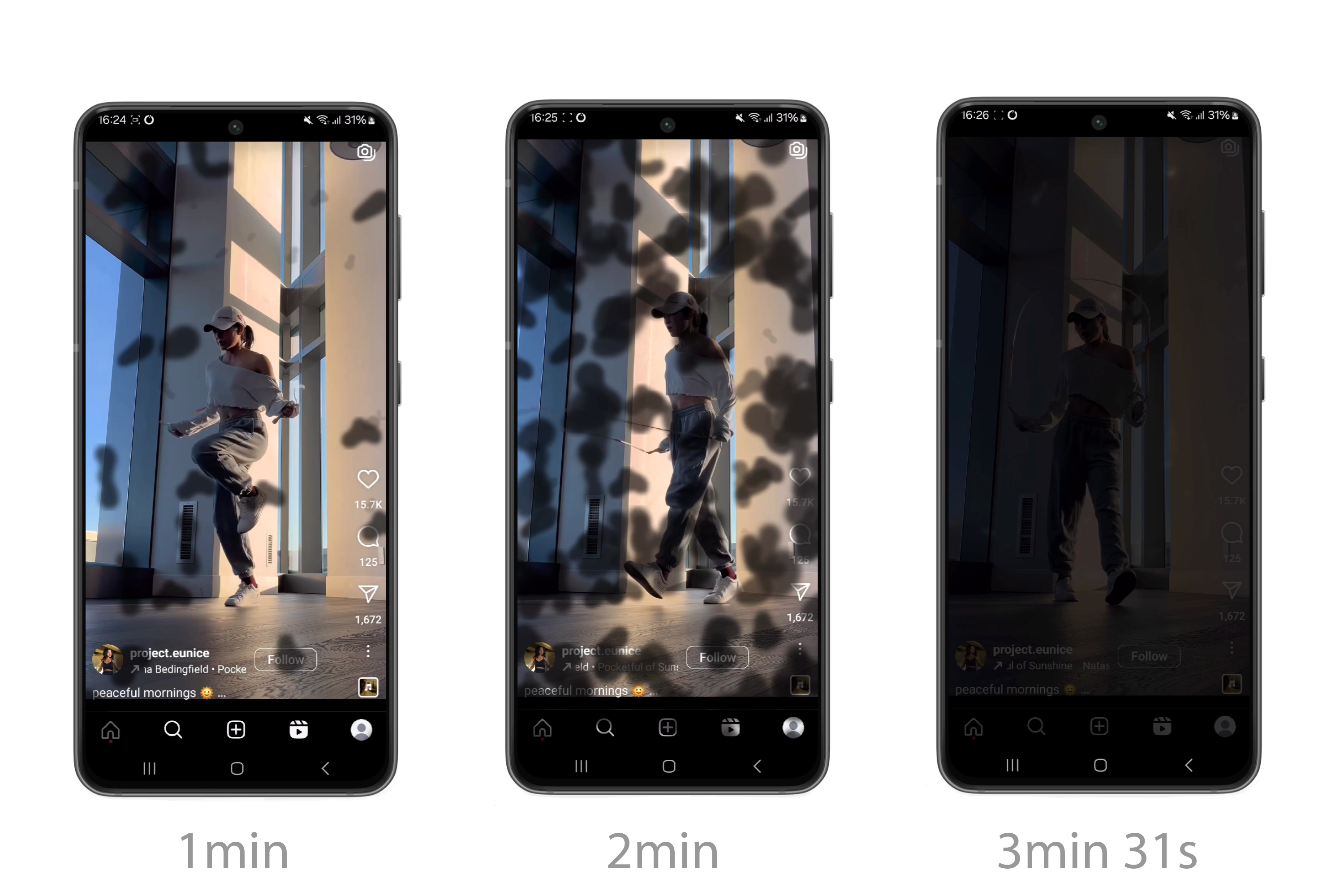}
        \caption{Visual Intervention}\label{fig:visual_intervention}
        \Description{Visual Intervention: Three smartphone screens show how black spots gradually obscure the video feed over time. At 1 minute, the screen is clear, at 2 minutes it is partially covered with black spots, and by 3 minutes 31 seconds, the video is fully obscured.}
    \end{subfigure}
        % \hspace{1.4cm}
    \begin{subfigure}[c]{0.38\linewidth}
        \includegraphics[height=3.8cm]{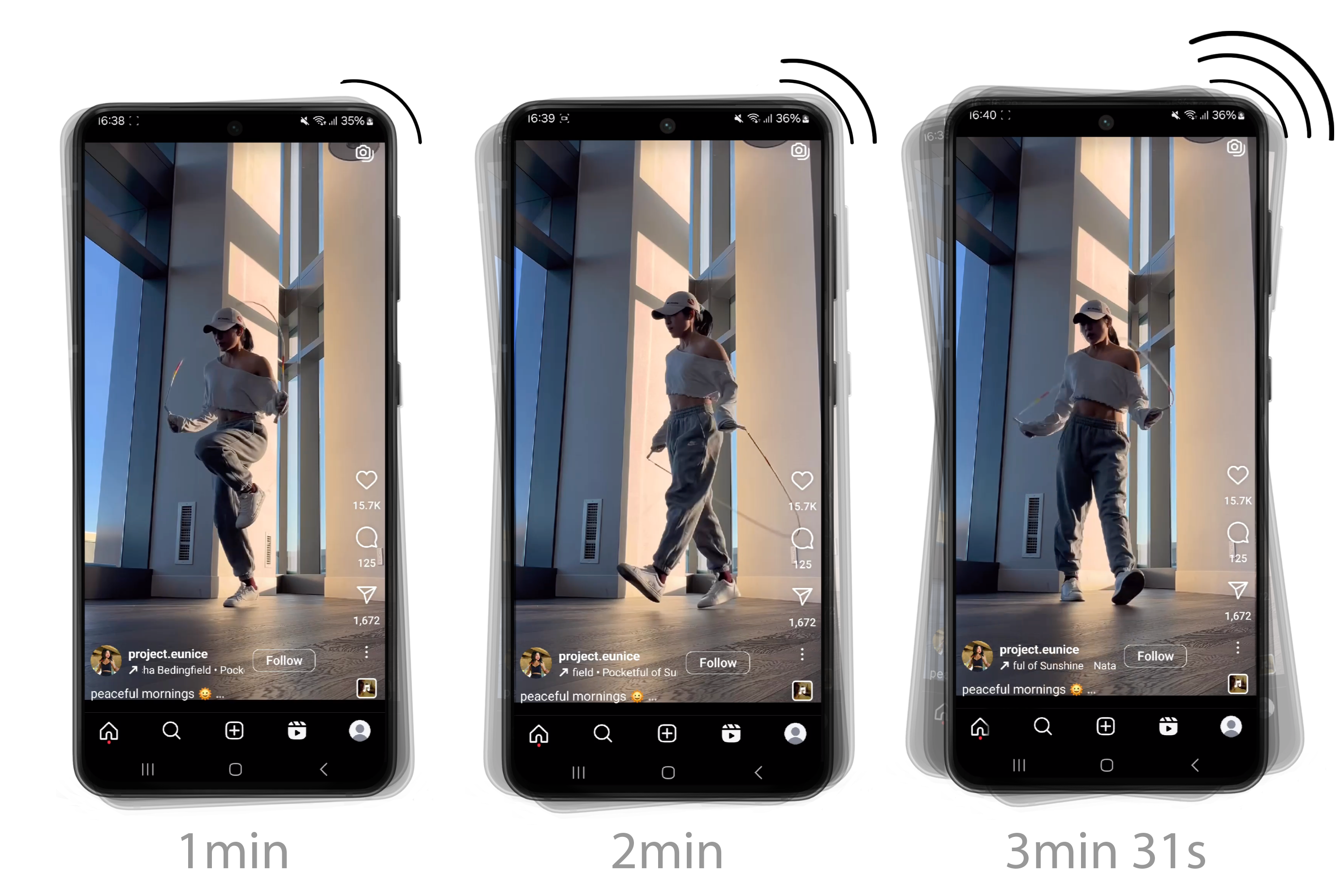}
        \caption{Haptic Intervention}\label{fig:haptic_intervention}
        \Description{Haptic Intervention: Three smartphone screens depict increasing vibration feedback while scrolling. At 1 minute, no vibration is shown, at 2 minutes, a single vibration icon appears above the phone, and at 3 minutes and 31 seconds, multiple vibration wave icons indicate stronger intensity.}
    \end{subfigure}
    % \hspace{1.4cm}
  
   \caption{Visualization of the three interventions that were used in the user study. The Visual and Haptic Intervention increases friction intensity over time, maximizing at 3~min 31~s.}~\label{fig:interventions}
   \Description{The figure shows three smartphone-based interventions to reduce infinite scrolling}
\end{figure*}

The baseline intervention replicated prior work~\cite{meinhardt_scrolling_2025} and existing screen-time reminders used on TikTok~\cite{TikTokScreenTime2025} and Instagram~\cite{InstagramTimeLimit2025}. The intervention consisted of a single pop-up overlay ($310 \times 400$\,dp), presented once per scrolling session, that informed users it was time to close the SoMe app and take a break from scrolling. Users could dismiss this message by tapping the “Dismiss” button, allowing them to continue scrolling (see \autoref{fig:baseline_intervention}). We compared this baseline intervention with two novel interventions that gradually intensify based on haptic and visual modality. Drawing on the principle of gradual stimuli for behavior changes~\cite{kincaid_gradual_2023}, these interventions subtly disrupt the user experience of infinite scrolling over time. By gradually increasing the cost of infinite scrolling through visual and haptic interference, we aimed to encourage disengagement without abrupt disruption. This approach aligns with design friction~\cite{cox_design_2016, ruiz_design_2024}, which advocates for thoughtful interaction interruptions to support digital self-regulation. Similar to the baseline, the gradual interventions were triggered after 15\,min of uninterrupted infinite scrolling and reached maximum intensity
after 3\,min~31\,s. This duration was grounded in related work~\cite{meinhardt_scrolling_2025}, where users took an average of 3\,min~31\,s to stop infinite scrolling after an intervention occurred. %We adopted this response time as the duration for our gradual interventions. 

For the \textbf{visual intervention}, 12 different semi-transparent black spot images ($37 \times 41$\,px to $84 \times 83$\,px) were placed at random positions across the phone's screen, inspired by the visual impairment of diabetic retinopathy~\cite{nei_diabetic_nodate}. A total of 3{,}000 spots were added over the time window, with 90\% of all spots appearing in the final 30\,s. Each spot ran an independent 15\,s fade-in animation, scaling from 10\% to 100\% of its size while increasing from fully transparent to near-opaque. 
For the \textbf{haptic intervention}, we build on the approach from \citet{okeke2018good}. A recurring 500\,ms vibration pulse started at amplitude 30 (of 255)~\cite{android:vibrationeffect.createOneShot}, incrementing by 3 per pulse until reaching maximum. The initial pause between pulses was 5\,s, decreasing by 70\,ms after each pulse until reaching 0\,ms. The 5\,s initial pause follows \citet{Pielot2014Vibration}, who showed that this pause between vibrations prevents users from quickly adapting to repetitive haptic feedback.

Our design choices for the gradual interventions aimed to compare how contextual factors and individual traits moderate the effectiveness of different friction modalities suitable for in-field deployment. Semi-transparent visual occlusion was chosen to introduce perceptual friction while keeping content visible. Haptic friction was chosen as a non-visual modality that conveys friction with minimal reliance on sound, which may be perceived as intrusive in public or quiet environments and, therefore, less suitable for passive smartphone use~\cite{Pielot2014Vibration}. 
% Although we evaluated the baseline alongside the gradual interventions, it represents current industry practice in digital well-being rather than a sensory-friction modality~\cite{TikTokScreenTime2025, InstagramTimeLimit2025}. The pop-up is easy to interpret and can be dismissed immediately, unlike the gradual interventions that increase over time and cannot be reversed without ending the scrolling session. 
% We include the baseline to anchor the study in a realistic intervention design that users commonly encounter on SoMe platforms. 
% This design was intended to support the core focus of the study, which is to explore how users with different personalities and in different contexts respond to different kinds of interventions.

\subsection{Procedure}\label{sec:procedure}
Recruitment and screening were conducted on \href{https://www.prolific.com}{Prolific} with participants from the United States and the United Kingdom. Eligibility was restricted to individuals who owned a device with Android 10 or later to leverage platform-specific permission handling introduced with Android 10. At the time of the study, this version was installed on over 87\% of Android devices in both countries~\cite{AndroidVersionUSA}. After accepting the Prolific study, participants were redirected to an \href{https://www.limesurvey.org/}{LimeSurvey} instance where they completed the registration survey, including the trait questionnaires assessing \textit{anxiety}, \textit{self-control}, \textit{FOMO}, and \textit{impulsivity}. To ensure attentiveness, we included the attention check: ``I breathe more than once a day.''
Upon completion, they received a tutorial video and a QR code (for desktop users) or a direct download link (for mobile users) to install the study application on their primary Android smartphone. After installation, the app displayed a consent form, which participants were asked to review and accept. Participants who installed the application and completed the full 7-day duration (regardless of whether they provided data) were compensated £1.22/\$1.55. No dedicated fraud screening was implemented, but automated participation was unlikely as in-app questionnaires only appeared after 15 minutes of natural scrolling on participants' phones, making their timing unpredictable in daily use.

The study protocol, including the consent procedure, received formal approval from the university's Ethics Committee. Particularly, those ensuring participant anonymity were strictly adhered to. Following consent, participants were guided through enabling the necessary permissions. The app then prompted users to enter their age and select their gender. Once these demographic items were completed, the app activated its main tracking function. 
% This was indicated by a persistent icon in the system’s status bar, as mandated by Android for all foreground services~\footnote{\url{https://developer.android.com/develop/background-work/services/foreground-services}, accessed: July 14, 2025}. The icon remained visible for the entire duration of the study. Because of this constant presence, it is assumed that participants are habituated to the icon, minimizing any influence on their typical scrolling behavior.
At the end of the 7-day period, the app notified participants that the study had concluded and that they could uninstall the application. We then compensated the participants according to the number of questionnaires filled out (£0.50/\$0.60 per questionnaire).

\subsection{Measurements}\label{sec:meaurements}
% This section describes the specific items used to assess subjective and objective intervention effectiveness, contextual factors, and personality traits throughout the study. 
To minimize participant burden and reduce the risk of survey fatigue, we primarily relied on short, single-item measures. Although single-item assessments may introduce some measurement error~\cite{Dejonckheere.2022}, their use is widely accepted and validated in SoMe research, offering a practical trade-off between accuracy and user engagement~\cite{beyens2020effect, RixenIS.2023, Bayer.2018, Chang.2015, meinhardt_scrolling_2025}. To ensure data quality, we included random attention checks into the questionnaires. A full list of the items used in the study is provided in \autoref{app:usedquestionaires}.

\subsubsection{Intervention Effectiveness}\label{sec:intervention_effectiveness}
We assessed the effectiveness of the three interventions using both objective and subjective measures to capture behavioral change and user experience.

\textbf{Objective effectiveness} was measured using the time it took participants to disengage from infinite scrolling after the intervention occurred. This \textit{responsiveness} was measured in seconds. To address the right-skewness in the \textit{responsiveness}, we applied a logarithmic transformation~\cite{draper1998applied}. Subsequently, to calculate the objective effectiveness score, we normalized the data to a 0–1 scale, using min–max normalization based on the observed range, and reversed the outcome so that higher scores indicate faster disengagement.

\textbf{For the subjective effectiveness}, participants rated the interventions based on five scales after they stopped infinite scrolling. These measures were derived from (1) \citet{kai_internal}, (2) \citet{Venkatesh2000}, and (3) \citet{meinhardt_scrolling_2025}. From \citet{kai_internal}, we used the items to measure \textit{goal alignment}, \textit{satisfaction}, and \textit{agency}; all three items were rated on a 7-point Likert scale. The wording of the items was slightly modified to ensure they referred directly to the intervention experience. For example, to assess perceived agency, participants were asked: ``\textit{For this intervention, how much did you feel out of or in control?}''. In addition to these three measures, we included the perceived \textit{usefulness} item drawn from the extended Technology Acceptance Model~\cite{Venkatesh2000}. Perceived usefulness is a determinant for the intention to use~\cite{Venkatesh2000}; therefore, we believe it is an ideal sub-indicator for the subjective effectiveness of an intervention. Usefulness was measured by the single-item ``\textit{I find the intervention to be useful in my current situation}'' rated on a 7-point Likert scale. Lastly, to capture potential negative reactions, we measured \textit{Reactance} using the threat subscale of the Reactance Scale for HCI (RSHCI)~\cite{Ehrenbrink.2020}, following the approach of \citet{meinhardt_scrolling_2025}. Reactance refers to the resistance individuals feel when their freedom is restricted~\cite{Ehrenbrink.2020}. This resistance is rooted in psychological models~\cite{Dillard.2005}, implying that ``\textit{messages [interventions] designed with the objective of behavior change must necessarily (implicitly or explicitly) limit an audience’s freedom}''~\cite[p. 67]{Rains.2013}. Thus, it is particularly relevant here, as interventions may limit users' ability to continue scrolling. For the threat subscale in RSHCI, five items were used, such as ``\textit{I don’t want the intervention to tell me what to do}'', which were rated on a 5-point Likert scale (1= "strongly disagree", 5="strongly agree"). We reversed the rates for reactance and normalized each score to a 0-1 range, respectively, to each questionnaire scale's boundary. Subsequently, we combined the five subjective measures into a single weighted score for the subjective intervention effectiveness. Hence, we conducted an exploratory factor analysis~\cite{efa} using the minimum residual method, which supported a unidimensional structure. All items loaded substantially on the latent factor (threshold $>0.4$~\cite{raykov_introduction_2011}): Reactance (0.49), Goal Alignment (0.89), Satisfaction (0.88), Agency (0.51), and Usefulness (0.85). Model fit indices acceptable fit (RMSR = 0.07), with strong reliability statistics ($\omega = 0.90$). Following established practice~\cite{distefano_understanding_nodate}, we used the normalized factor loadings as weights to calculate the combined subjective effectiveness score.

\subsubsection{Contextual Factors}\label{sec:measure_context_factors}
To determine the effects of context on intervention effectiveness, we included seven contextual factors. For \textit{current activity}, we applied the interval scale developed by \citet{Samdahl.1991}, ranging from -3 (“definitely leisure”) to +3 (“definitely not leisure”). The \textit{social situation} was measured using a question adopted from \citet{Akpinar.2023}, asking participants: “Which one of these best describes people around you?” with two response options: alone and with acquaintances (e.g., friends, family, colleagues). To determine whether participants were \textit{at home}, we asked a simple yes/no question: “Are you currently at home?” adopted from \citet{meinhardt_scrolling_2025}. We determined \textit{multitasking} by asking, “Did you do anything else besides being on [App Name]?” which participants could also answer with yes or no~\cite{meinhardt_scrolling_2025}. For evaluating participants’ \textit{valence}, we used the Self-Assessment Manikin (SAM) scale~\cite{Bradley.1994}, as previously employed by \citet{RixenIS.2023}. Sleepiness was measured using the Karolinska Sleepiness Scale (KSS)~\cite{Shahid.2012b}, which ranges from 1 (“extremely alert”) to 9 (“extremely sleepy”). Lastly, stress was determined by the stress numerical rating scale~\cite{Karvounides2016} on an 11-point Likert-type scale, ranging from 0 (“no stress”) to 10 (“worst stress possible”). Using the fixed boundaries of each questionnaire, we normalized all contextual factors to a range from 0 to 1.

\subsubsection{Individual Traits}
We measured \textit{impulsivity} using the 15-item short version of the Barratt Impulsivity Scale~\cite{Spinella2007}. This scale assesses the subcategories of non-planning, motor impulsivity, and attentional impulsivity on a 4-point Likert scale. For \textit{self-control}, we used the 13-item Self-Control Scale developed by \citet{tangney_high_2004}. It is rated on a 5-point Likert scale. For \textit{anxiety}, we used the short version of the Spielberger Trait Anxiety Inventory, which consists of 5 items~\cite{Zsido2020}, using a 4-point Likert scale ranging from 1=almost never to 4=almost always. \textit{FOMO} was assessed via the trait part of the Trait-State FOMO scale, which uses 5 items on a 5-point Likert scale~\cite{Wegmann2017}. Using the fixed boundaries of each questionnaire, we normalized all individual traits to a range from 0 to 1.

\subsection{Participants and Power Simulation}\label{sec:power_simulation}
Only participants who experienced all three intervention types were included in the final analysis to maintain the integrity of our within-subject design. To determine the required sample size for achieving 80\% statistical power~\cite{cohen_power_1992} for our subsequent analysis (see \autoref{sec:baysian_model}), we followed the simulation-based approach proposed by \citet{Kumle.2021}. After collecting data from an initial batch of $n=30$ participants ($M=34.33$, $SD=10.76$ years; 11 female, 18 male, one non-binary, 0 prefer not to say), (334 data points), we attempted to run the power simulation using the full linear mixed model (LMM) with all 11 factors (see \autoref{sec:peronality_contextual_factors}). However, this model proved computationally infeasible to simulate for power, even after one month of runtime.

To address this, we reduced complexity by identifying the most important factors. Hence, we fit the full LMM to the pilot data and inspected the $t$-values of all interaction terms involving \textit{Intervention Type}. Following the threshold of $|t| \geq 2$~\cite{Kumle.2021}, we identified \textit{Impulsivity ($t = -2.69$)}, \textit{Anxiety ($t = 3.51$)}, and \textit{FOMO ($t = 2.22$)} as the factors with sufficiently strong interaction effects. The corresponding $t$-values are reported in \autoref{tab:xgb-feature-importance}. Based on these results, we simplified the model for the power simulation. To reduce computational effort in simulating two models for subjective and objective intervention effectiveness, we combined both scales, leading to the following model:
\[
\text{Inter. Effectiveness (combined)} \sim \text{Inter. Type} \times (\text{Impulsivity} + \text{Anxiety} + \text{FOMO}) + (1 \mid \text{Participant})
\]

\noindent Using the \texttt{mixedpower} package~\cite{Kumle.2021}, we ran simulations with the simplified model for increasing sample sizes ranging from 30 to 150 in steps of 30, with 1000 simulations per sample size and a critical value of 2 (approximating a two-tailed $\alpha$ level of .05~\cite{Kumle.2021}).
The results of the simulation (see \autoref{app:power_sim}) showed that the interaction between \textit{Intervention Type × Impulsivity} reached more than 80\% power at $n=30$. \textit{Intervention Type × FOMO} reached 80\% power by $n=60$. \textit{Intervention Type × Anxiety} reached 80\% power by $n=90$. Therefore, a sample size of at least $n=90$ is sufficient to detect the interaction effects with acceptable power~\cite{cohen_power_1992}. 

Hence, participant recruitment proceeded in batches of approximately 30 participants over a four-month period. 792 individuals completed the registration phase. Of these, 54 were excluded for failing attention checks in the individual traits questionnaires, and 524 were asked to return their submissions because they did not install the application. Participants who opted out often mentioned download difficulties or the burden of a 7-day study. In addition, we believe this dropout range can largely be attributed to the low initial effort required for registration. This left 214 participants who successfully completed the 7-day app-based study and received compensation for their participation.
After filtering out participants who did not experience all three intervention types (see above), the final dataset included $N=104$ participants ($M=36.95$, $SD=10.68$ years; 44 female, 58 male, 2 non-binary, 0 preferred not to say). Of the 110 excluded participants, 64 experienced two or fewer interventions in total, suggesting they either lost interest during the study period or did not exceed the 15-minute scrolling threshold frequently enough to trigger all conditions. The remaining 46 participants experienced three or more interventions overall, but due to the randomized assignment order, they were not exposed to all three types. To assess whether this filtering decision influenced the results, we conducted sensitivity analyses with two relaxed inclusion criteria ($N=150$ and $N=214$; see \autoref{app:sensitivity}).

Of the included participants, an additional 49 data points were excluded due to failed attention checks in the in-app surveys, and 23 incomplete entries were removed as they were likely due to technical errors. Subsequently, we applied the z-score method to detect outliers across all six intervention effectiveness measures. Data points exceeding $\pm 3$ SD were excluded, which resulted in removing 24 observations for \textit{responsiveness}. After this step, 1,294 data points from 104 participants remained for the final analysis ($M = 12.40$, $SD = 9.36$, $Md = 10$ interventions per participant), distributed across platforms as follows: 41.27\% TikTok, 22.91\% Instagram, 17.91\% Facebook, and 17.91\% YouTube Shorts. Each intervention was distributed nearly equally across participants, with a median of three occurrences per participant for all three conditions (Baseline: $M = 4.26$, $SD = 3.86$; Visual: $M = 4.23$, $SD = 3.52$; Haptic: $M = 4.03$, $SD = 3.35$). Further descriptive statistics are provided in \autoref{app:descriptive}.

\subsection{Results}
% The goal of our analysis was to understand how different types of interventions influence disengagement from infinite scrolling, both in terms of objective responsiveness and subjective ratings, and how this effectiveness is moderated by contextual factors and individual differences.
Initially, we conducted a survival analysis to examine disengagement behavior over time while accounting for censored observations. Second, we examined how participants’ experience of the interventions changes as their friction intensifies and when the gradual intervention becomes too intense and thus less effective (RQ1). Finally, we present two Bayesian mixed-effects models, which test how contextual factors and individual traits moderate subjective and objective intervention effectiveness (RQ2). 

\subsubsection{Disengagement from Infinite Scrolling: Descriptive and Survival Analysis}
Participants disengaged fastest with the baseline intervention ($Md = 7s$), followed by the haptic gradual intervention ($Md = 28s$), and slowest with the visual gradual intervention ($Md = 56s$). Further, the baseline exhibited lower variability ($IQR = 45s$) compared to the gradual interventions (visual: $IQR = 1min52s$; haptic: $IQR = 1min21s$).
\begin{figure}[ht!]
    \centering
    \includegraphics[width=0.6\linewidth]{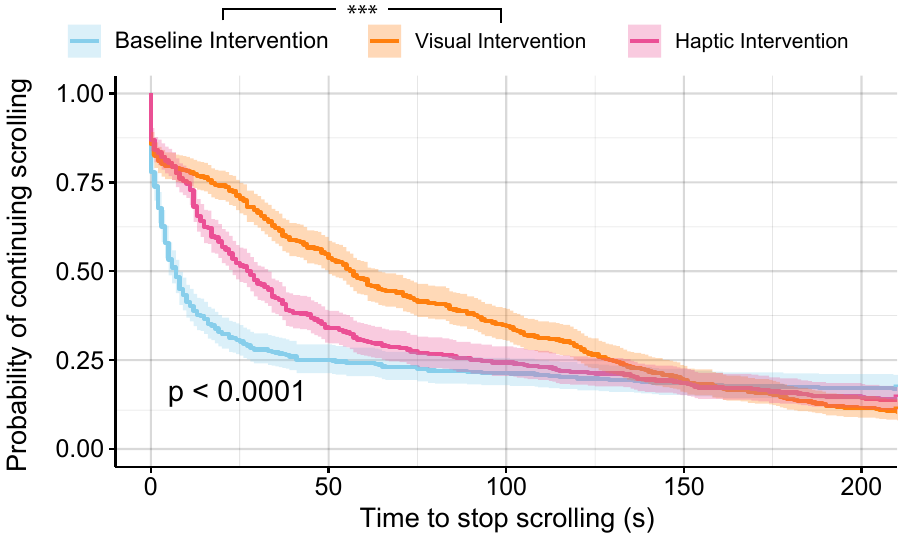}
    \caption{Kaplan-Meier curves show the probability of the participants continuing to scroll after an intervention occurred.}
    \label{fig:km_plot}
    \Description{The figure shows a survival curve of the probability of continuing to scroll over time for three interventions. The x-axis represents time to stop scrolling in seconds (0-200), and the y-axis shows the probability of continuing scrolling. The baseline intervention (blue) shows the fastest decline, indicating users stopped scrolling earlier. The haptic intervention (pink) also reduced scrolling but less strongly than baseline. The visual intervention (orange) shows the slowest decline, with users continuing to scroll longer. A statistical comparison indicates a significant difference between baseline and visual intervention.}
\end{figure}
As our data captured disengagement as a time-to-event process, we analyzed responsiveness using survival analysis. Survival analysis is well-suited here because it accounts for censoring~\cite{Kaplan01061958}, that is, cases where participants continued scrolling beyond the observation window. We defined this window as the point of maximum intervention intensity (3~min 31~s). Responsiveness values below this cutoff were coded as events (scrolling stopped), values exceeding it were treated as censored. We then fitted a Cox proportional hazards model, a regression approach within survival analysis that estimates how predictors (here, intervention type) affect the hazard rate (HR), which is the instantaneous probability of stopping scrolling at a given time~\cite{cox_regression_1972}. To account for repeated observations per participant, clustering was included at the participant level. The model indicated significant overall differences between interventions, $\chi^2$(2) = 22.13, $p < .001$.
Pairwise comparisons using estimated marginal means (on the log-hazard scale) showed that the visual gradual intervention led to significantly slower disengagement than the baseline pop-up, $HR = 0.71$, 95\% CI [0.59, 0.85], $p < .001$. The haptic gradual intervention also reduced disengagement relative to the pop-up ($HR = 0.80$, 95\% CI [0.65, 0.99]), but this effect was weaker and only marginally reliable after Tukey correction ($p = .043$, \padj{0.11}). The visual and haptic interventions did not significantly differ from each other, $HR = 0.88$, 95\% CI [0.72, 1.08], \padj{0.21}. Kaplan-Meier curves visualize these effects (see \autoref{fig:km_plot}).

\paragraph{Re-engagement Analysis} \label{sec:reengagement}
We analyzed re-engagement patterns between consecutive scrolling sessions to assess whether the interventions influenced users to resume scrolling directly after disengagement. Hence, we measured the break time from the disengagement of scrolling due to an intervention to when users began scrolling again in the next session. It is essential to note that only sessions in which participants exceeded the 15-minute threshold (triggering an intervention) were recorded; sessions in which users naturally disengaged before reaching this threshold generated no data points. Across 1,190 consecutive intervention-triggered sessions, 4.87\%  re-engaged within one minute, 11.5\% within 3 minutes, 13.70\% within five minutes, and 17.5\% within ten minutes. When broken down by intervention type, rapid re-engagement (within 5 minutes) rates were similar: Baseline (12.5\%), visual (14.2\%), and haptic (14.4\%). The median break time was 3.4~h overall, with slight variations across interventions: Baseline ($Md = 4.3$~h, $IQR = 18.5$~h), visual ($Md = 3.3$~h, $IQR = 13.7$~h), and haptic ($Md = 3.2$~h, $IQR = 14.6$~h). See \autoref{app:descriptive}, \autoref{fig:reengagement_histogram} for the histogram.

\subsubsection{Subjective Effectiveness of the Gradual Interventions Over Time} 
\begin{figure}[ht!]
    \centering
    \includegraphics[width=0.93\linewidth]{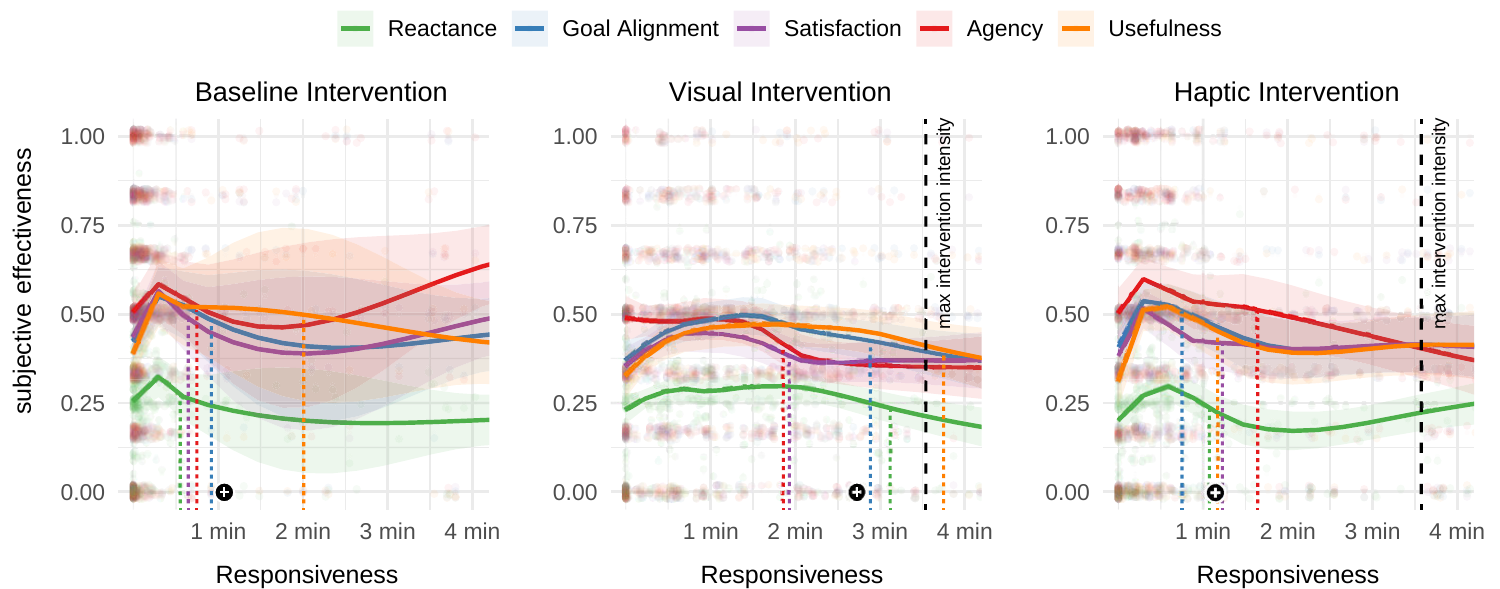}
    \caption{Relationship between responsiveness and interventions' subjective ratings. Dots represent individual data points, and lines show LOESS-smoothed ($span=0.75$, \textit{default}). The x-axis shows the time participants needed to respond to the intervention by stopping infinite scrolling. After 3~min 31~s, the gradual interventions reached their maximum in intensity. The colored dashed lines mark the points at which each subjective rating dropped below at least one point on the respective Likert scale. The black dot, with the white cross, indicates the weighted mean where the subjective effectiveness drops.}
    \Description{The figure shows three line graphs for the subjective effectiveness ratings for baseline, visual, and haptic interventions over time. The x-axis represents responsiveness, defined as the time until participants responded to the intervention, and the y-axis shows subjective effectiveness on a scale from 0 to 1. Each graph includes LOESS-smoothed lines for five measures: reactance (green), goal alignment (blue), satisfaction (red), agency (purple), and usefulness (orange). In the baseline intervention, satisfaction and agency begin high but decrease as time progresses. In the visual intervention, ratings remain relatively stable, with satisfaction and agency consistently higher than reactance. In the haptic intervention, satisfaction and agency also start high but decline more noticeably, while reactance increases. Colored dashed lines indicate when ratings dropped by at least one Likert point, and each graph includes a black dot with a white cross showing the mean point at which subjective effectiveness declined.}
    \label{fig:time_plot}
\end{figure}

RQ1 examines how participants’ perceptions of intervention effectiveness change over time as the gradual interventions increase friction. To explore this, we analyzed how the subjective effectiveness developed in relation to the time it took participants to stop infinite scrolling
% . This approach allowed us to capture not only whether the interventions were objectively effective, but also when participants began to perceive them over time
(see \autoref{fig:time_plot}). 
The baseline intervention showed a peak in all ratings around 15~s after the pop-up appeared, but these ratings dropped and fluctuated afterwards (high standard error), suggesting a fading impact if no immediate action was taken. The haptic gradual intervention also peaked at around 15~s, but ratings then remained comparatively stable as the vibration intensity increased. The visual gradual intervention showed a different pattern: ratings stayed steady throughout without an initial peak, maintaining a level similar to the haptic intervention but lacking the early boost observed in the other two conditions. 

To identify when interventions began to lose subjective effectiveness, we defined a \textit{tipping point} as the moment when ratings dropped by at least one point on their respective Likert scales. Since measures were normalized, this corresponded to a drop of at least $1/7$ for seven-point scales and $1/5$ for the five-point scale of \textit{reactance} (reversed). One Likert step represents the minimal perceptible change a participant could report. The weighted means (using the factor loadings from our exploratory factor analysis (see \autoref{sec:intervention_effectiveness}) of the five subjective ratings were calculated. These tipping points are visualized in \autoref{fig:time_plot} and summarized in \autoref{tab:intervention_comparison}.  

\begin{table}[h!]
\footnotesize
    \caption{Tipping points where the subjective effectiveness drops by at least one step in the respective Likert scale}
    \label{tab:intervention_comparison}
    \centering
    \begin{tabular}{lccc}
    \toprule
    \textbf{Subjective Measure} & \textbf{Baseline Inter.} & \textbf{Visual Inter.} & \textbf{Haptic Inter.} \\
    \midrule
    Agency & 45~s & 111~s & 97~s \\
    Goal Alignment & 57~s & 173~s & 71~s \\
    Satisfaction & 39~s & 115~s & 42~s \\
    Reactance (reversed) & 31~s & 185~s & 64~s \\
    Usefulness & 120~s & 221~s & 69~s \\
    \midrule
    \textit{Weighted Mean} & \textit{62.21~s} & \textit{163.06~s} & \textit{66.19~s} \\
    \bottomrule
    \end{tabular}
\end{table}

\noindent The results indicate that ratings for the haptic intervention declined earlier than those for the visual intervention. For the haptic intervention, the subjective effectiveness dropped at 1~min 6~s, while the visual intervention’s subjective effectiveness dropped at more than double the time (2~min 43~s). The baseline's subjective effectiveness dropped the earliest at 62.21~s. However, this timing should be interpreted cautiously due to the large standard error after 15~s.  

\subsubsection{Correlation Analysis Between Individual Traits}\label{sec:construct_overlap}
We computed pairwise Pearson correlations at the participant level to assess overlap among the four individual traits. Impulsivity and self-control showed the strongest negative correlation ($r = -.72$, $p < .001$). Strong correlations also emerged between self-control and FOMO ($r = -.62$, $p < .001$) and between anxiety and FOMO ($r = .60$, $p < .001$). The remaining pairs showed moderate correlations: self-control and anxiety ($r = -.54$, $p < .001$), impulsivity and FOMO ($r = .45$, $p < .001$), and impulsivity and anxiety ($r = .41$, $p < .001$).

\subsubsection{Bayesian Mixed-Effect-Models} \label{sec:baysian_model}
RQ2 addresses how contextual factors (within-subject) and individual traits (between-subject) moderate the effectiveness of interventions. We distinguish between subjective (user experience) and objective effectiveness (behavior change). 
% This dual perspective reflects both how quickly participants disengaged from infinite scrolling and how they perceived the intervention. 
Although our central focus is on moderation effects, we first report descriptive results to provide an overall orientation. Across all participants, the baseline intervention achieved the highest subjective and objective effectiveness (subjective: $M = 0.48$, $SD = 0.23$; objective: $M = 0.65$, $SD = 0.30$). The haptic gradual intervention followed (subjective: $M = 0.42$, $SD = 0.23$; objective: $M = 0.55$, $SD = 0.25$), while the visual gradual intervention was rated lowest on both dimensions (subjective: $M = 0.39$, $SD = 0.21$; objective: $M = 0.51$, $SD = 0.26$).

Unlike classical models, Bayesian methods allow quantifying the significance and existence of effects, making them particularly suitable for small-sample and novel HCI contexts~\cite{kay_researcher-centered_2016, schad_toward_2021}. Although we carefully simulated the sample size prior to our user study (see \autoref{sec:power_simulation}), using Bayesian statistics enables us to perform a more robust analysis. While still uncommon in HCI research~\cite{kay_researcher-centered_2016}, Bayesian methods are increasingly applied in behavioral sciences~\cite{van_de_schoot_systematic_2017}.  

Initially, we tested for normality of our data. A Shapiro–Wilk test indicated non-normality for subjective ($W = 0.98$, $p < .001$) and objective effectiveness ($W = 0.95$, $p < .001$). Therefore, we specified Student-$t$ distributions. %As \citet{kruschke_doing_2015} notes, 
The $t$ distribution is useful for Bayesian models ``\textit{when data appear to have outliers beyond what would be accommodated by a normal distribution}''~\cite[p.~458]{kruschke_doing_2015}. The model was fitted using Markov-chain Monte-Carlo (MCMC) sampling (4 chains, 11{,}000 iterations per chain, 1{,}000 warmup). As suggested by \citet{lemoine_moving_2019}, we used $B {\sim} \mathcal{N}(0, 1)$ as a weakly informative prior.
Such priors act as a form of regularization, constraining estimates to plausible ranges and improving robustness, especially in complex models like ours, where even a reasonably large sample size (see \autoref{sec:power_simulation}) may not fully counteract the risk of non-convergence, when not using priors~\cite{schad_toward_2021}.
 The model structure was:
{\small
\[
\text{Subj./Obj. Effectiveness} \sim \text{Inter. Type} \times (\text{Within-Factors} + \text{Between-Factors}) + (1 + \text{Inter. Type} + \text{Within-Factors} \mid \text{Participant})
\]
}

\noindent The within-subject factors were valence, stress, sleepiness, multitasking, at home, current activity, and social situation; between-subject factors included impulsivity, anxiety, FOMO, stress, and self-control. The model for the subjective effectiveness showed substantial explanatory power ($Conditional~R^2 = 0.80$ (95\% CrI $[0.79, 0.81]$), $Marginal~R^2 = 0.28$, (95\% CrI $[0.21, 0.35]$) while the low model for the objective effectiveness was lower powered ($Conditional~R^2 = 0.45$ (95\% CrI $[0.40, 0.50]$), $Marginal~R^2 = 0.12$, (95\% CrI $[0.09, 0.17]$). All potential scale reduction factors for both models' parameters were below  1.01, indicating good convergence~\cite{Brooks01121998}. Trace plots of the MCMC permutations were inspected for divergent transitions. In addition, the effective sample sizes (ESS) exceeded $12{,}100$ for all parameters, well above the $ESS > 10{,}000$ threshold suggested by \citet{kruschke_doing_2015} for obtaining reasonably stable estimates.

\citet{makowski_indices_2019} describes an effect as possible existing when their $pd > 95\%$. Further, they mention that an effect is considered as probably significant when less than 2.5\% of the 89\% Highest Density Interval (HDI) lies inside the Region of Practical Equivalence (ROPE), indicating a meaningful deviation from the null region. According to \citet{kruschke_bayesian_2018}, the ROPE range can be defined as a range corresponding to \(\pm0.1\) of the standard deviation of the target variable~\cite{cohen_statistical_1988}. We calculated the ROPE for the subjective effectiveness to be $[-0.02, 0.02]$ and for the objective effectiveness to be $[-0.03, 0.03]$. In the following, we report only the interaction terms that involve intervention type, as these directly address our RQ2. Several substantial main effects that replicate results from prior work~\cite{meinhardt_scrolling_2025, RixenIS.2023} were also identified, such as on valence and sleepiness. Additionally, impulsivity showed a significant main effect on subjective effectiveness (see \autoref{app:bayesian_model}). However, they fall outside the scope of RQ2 and are therefore not detailed here. 

We used the baseline intervention (Pop-up) as a reference condition, comparing it to the haptic and visual gradual intervention. Each effect is summarized by its posterior median, 95\% CrI, probability of direction (pd), and percentage of the posterior distribution within ROPE. We followed the reporting guidelines of \citet{kruschke_bayesian_2021} and \citet{makowski_indices_2019}.

\noindent \paragraph{\textbf{Subjective Effectiveness}}

\begin{figure}[ht]
    \centering
    \includegraphics[width=0.85\linewidth]{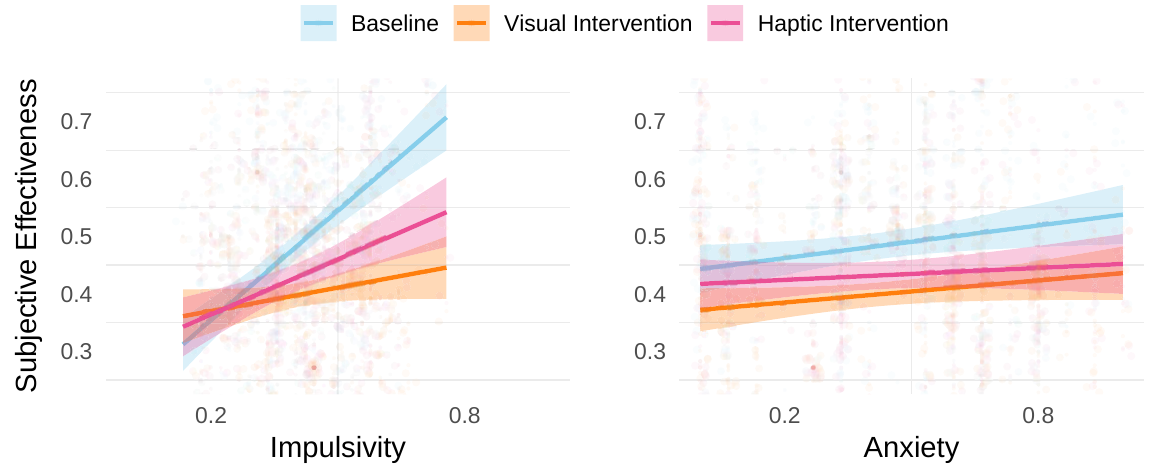}
    \caption{Interaction effects for impulsivity, and anxiety with intervention type on \textbf{subjective effectiveness}}
    \Description{The figure shows two line graphs of interaction effects between intervention type and individual traits on subjective effectiveness. The left graph plots impulsivity on the x-axis and subjective effectiveness on the y-axis. All three intervention types show increasing effectiveness with higher impulsivity, with the baseline condition showing the steepest rise. The right graph plots anxiety on the x-axis and subjective effectiveness on the y-axis. Effectiveness also increases with higher anxiety, but the slopes are shallower. Across both graphs, the baseline intervention yields the highest subjective effectiveness, followed by haptic, with visual being the lowest.}
    \label{fig:lineplots_interaction_effects_subjective}
\end{figure}

Participants with higher impulsivity rated the visual intervention as less effective than the baseline (see \autoref{fig:lineplots_interaction_effects_subjective} left). This interaction between \textit{Visual Intervention × Impulsivity}, observed on the subjective effectiveness as the weighted mean of the five subjective measures, probably exists ($pd=98.69\%$; Md = -0.29, 95\% CrI [-0.56, -0.03]) and is considered significant (0.00\% in ROPE).

We further found a trend for \textit{Visual Intervention × Anxiety} (Md = 0.12, 95\% CrI [0.00, 0.23], pd = 97.81\%, percentage in ROPE = 2.82\%). While the pd is $>$95\%, suggesting a possible existing effect, the percentage in ROPE shows undecided significance. Hence, the effect should be interpreted with caution (see \autoref{fig:lineplots_interaction_effects_subjective} right).

\noindent \paragraph{\textbf{Objective Effectiveness}}

\begin{figure}[ht]
    \centering
    \includegraphics[width=0.88\linewidth]{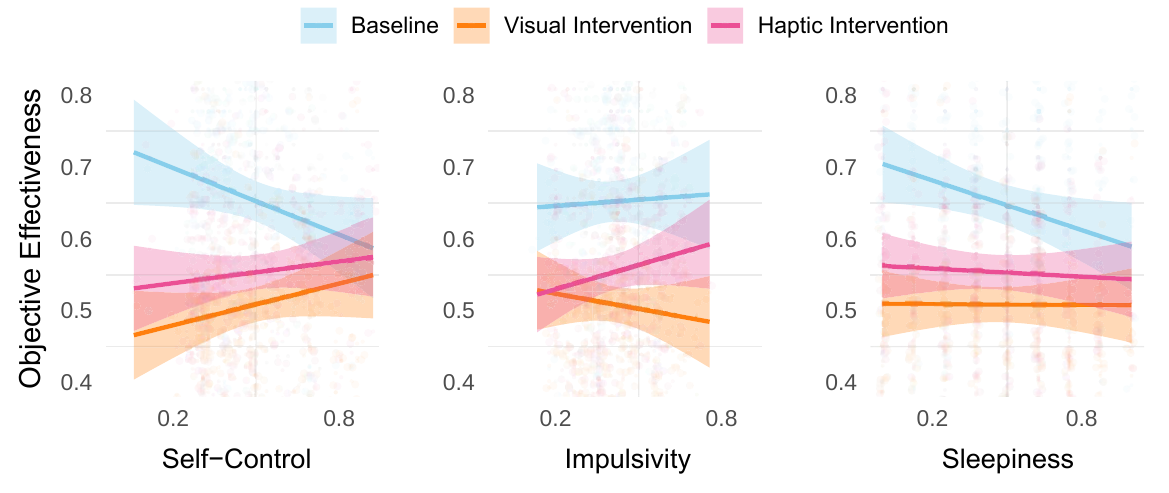}
    \caption{Interaction effects for self-control, impulsivity, and sleepiness with intervention type on \textbf{objective effectiveness}}
    \Description{The figure presents three line graphs showing interaction effects between intervention type and individual traits on objective effectiveness. The first graph plots self-control on the x-axis. Objective effectiveness decreases with higher self-control for the baseline condition (blue) but stays relatively stable for visual (orange) and haptic (pink) interventions. The second graph plots impulsivity. Objective effectiveness increases with impulsivity for the haptic intervention, decreases for the visual intervention, and rises slightly for the baseline condition. The third graph plots sleepiness. Effectiveness declines with increasing sleepiness in the baseline condition, while visual and haptic interventions remain relatively flat. Across all three traits, the baseline intervention generally shows higher effectiveness at low levels of self-control, impulsivity, and sleepiness, while differences between interventions become smaller at higher levels.}
    \label{fig:lineplots_interaction_effects_objective}
\end{figure}

Participants with low self-control yielded more objective effectiveness with the baseline than with the haptic intervention, while participants with high self-control yielded similar effectiveness with both intervention types (see \autoref{fig:lineplots_interaction_effects_objective} left). This interaction between \textit{Haptic Intervention × Self-Control} (Md = 0.58, 95\% CrI [0.18, 0.97], $pd$ = 99.7\%, 0\% in ROPE) probably exists and is significant.

Further, for participants with low impulsivity, the haptic intervention yielded lower objective effectiveness than the baseline, a difference that diminished with increasing impulsivity (see \autoref{fig:lineplots_interaction_effects_objective} center). This interaction between \textit{Haptic Intervention × Impulsivity} (Md = 0.42, 95\% CrI [-0.03, 0.86]) probably exists ($pd$ = 96.6\%) and is probably significant (1.91\% in ROPE).

Last, for sleepy participants, the objective effectiveness of the visual intervention was comparable to the baseline, whereas for more alert participants, the baseline intervention was more objectively effective (see \autoref{fig:lineplots_interaction_effects_objective} right). This interaction between \textit{Visual Intervention × Sleepiness} (Md = 0.13, 95\% CrI [0.01, 0.26], $pd$ = 98.0\%) likely exists, yet, as the percentage in ROPE is greater than 2.5\% (2.8\%), it has undecided significance and this trend should be handled with care.

\noindent \paragraph{\textbf{Other Effects}}
Several other interactions, including those with FOMO, or contextual factors such as multitasking, being at home, current activity, and social situation, showed uncertain existence (pd < 95\%) or undecided significance (ROPE > 2.5\%). This indicates that these factors are unlikely to moderate the relationship between intervention type and intervention effectiveness meaningfully. Full results of the Bayesian model are provided in \autoref{app:bayesian_model}. To assess the robustness of these findings, we conducted sensitivity analyses by re-calculating both models under two alternative participant filtering conditions ($N=150$ and $N=214$; see \autoref{app:sensitivity}). The \textit{Visual Intervention $\times$ Anxiety} (subjective effectiveness) and \textit{Haptic Intervention $\times$ Self-Control} (objective effectiveness) interactions remained significant across all filtering. The interaction between impulsivity and the intervention types attenuated when participants without complete within-person exposure were included.

\section{Discussion}
This study examined how the effectiveness of interventions for mitigating infinite scrolling on short video SoMe platforms is shaped by situational context and individual differences in the traits impulsivity, self-control, anxiety, and FOMO. We compared three designs: a baseline pop-up intervention, a gradually intensifying visual overlay, and a gradually intensifying haptic vibration. We assumed that no single intervention works best for all situations or all users. %An intervention might be highly effective in one context yet ineffective in another, or work well for individuals with certain personality traits but fail for others. 
% While prior work explored the importance of optimal timing for interventions~\cite{purohit_2019, orzikulova_time2stop_2024} and tested different intervention designs~\cite{mongeroffarello2019race, lu_interactout_2024}, these two research directions have typically been investigated separately. Our study integrates these two perspectives. % by investigating whether specific intervention types interact with particular contextual factors and personality traits. 
To answer our two RQs (see \autoref{sec:intro}), we conducted a 7-day field study ($N=104$) using a custom Android app to track scrolling on TikTok, Instagram, Facebook, and YouTube Shorts, with interventions triggering after 15\,min. %After 15 minutes of uninterrupted scrolling (in line with~\cite{meinhardt_scrolling_2025, RixenIS.2023, Terzimehic.2022b}), one of the interventions was triggered. 
During the study, upon stopping their scrolling activity, we assessed the intervention's effectiveness via objective and subjective measures and asked participants to report their current context (e.g., sleepiness, valence, etc.).

\subsection{User Experience as Intervention Intensity Increases (RQ1)}
To address user experience when the friction of interventions increases (see RQ1), we analyzed when subjective effectiveness declined, interpreting these declines as tipping points that may indicate the intervention was perceived as too intrusive. The baseline pop-up intervention worked well initially, but quickly lost subjective impact when participants did not disengage immediately and ignored it. The haptic gradual intervention extended subjective effectiveness somewhat beyond the baseline, but its ratings also declined after about 66\,s, likely because vibration intensity quickly became intrusive. In contrast, our gradual visual intervention maintained stable ratings for much longer, with declines occurring only after about 163\,s. While we can assume that it was better tolerated, it also yielded significantly slower behavioral responses than the baseline. These differences might be explained by the \textit{orienting response}~\cite{sokolov1960neuronal}, where unexpected signals in a different sensory modality automatically redirect attention. Because infinite scrolling is mainly a visually immersive activity, haptic vibrations triggered a stronger orienting response, making them harder to ignore but also more disruptive. In contrast, the increasing dots of the visual intervention blended into the visual stream of infinite scrolling, which may make them less intrusive and more naturally integrated, but also slower to prompt disengagement. 
% This interpretation aligns with research on cross-modal attention capture~\cite{spence_recent_2007}, which shows that task-irrelevant haptic signals can strongly disrupt ongoing visual or auditory tasks.

\subsection{Self-Regulation Moderates the Effectiveness of Interventions (RQ2)}
Subjective and objective effectiveness of the baseline pop-up for impulsivity diverged. Objectively, impulsivity did not affect disengagement time, but subjectively, highly impulsive participants rated the intervention as significantly more effective than less impulsive ones. One interpretation is that impulsive users may recognize their difficulty in self-regulating and appreciate the explicit cue of the pop-up as a supportive nudge on their behavior. Even if their stopping times are not substantially shorter, the explicit intervention may resonate with their self-perception of struggling to disengage, which makes the intervention subjectively experienced more effective. In contrast, less impulsive users, who rely more successfully on internal regulation, may perceive the same prompt as unnecessary, resulting in lower subjective ratings. However, subjective ratings were collected only from sessions in which participants stopped scrolling, so individual differences in response may also contribute to how the intervention was experienced. 

The divergence between subjective and objective effectiveness for impulsivity directly connects to the interaction effects with self-control on objective effectiveness: The baseline pop-up was objectively most effective for users with low self-control, but its impact declined for people with higher self-control. In contrast, the gradual interventions maintained relatively stable effectiveness across different levels of self-control. Taken together, these findings highlight that explicit interventions like our pop-up design are especially valuable for users who struggle with self-regulation (high impulsivity or low self-control). %Here, the choice of intervention highly matters. 
This interpretation aligns with \citet{mark_effects_2018} showing that people with lower self-control perceive intervention as more beneficial than those with high self-control. 
% More subtle interventions, such as visual gradual intervention, might be preferred.
While prior work has well established that traits like low self-control predict higher vulnerability to problematic SoMe use~\cite{simsir-gokalp_self-control_2024}, our results explicitly extend this relationship: individual traits do not merely predict problematic use, but \textit{moderate} the effectiveness of behavioral interventions. 
% Building on findings that people with lower self-control perceive interventions as more beneficial~\cite{mark_effects_2018}, our analysis reveals that they specifically rely on explicit friction to disengage. In contrast, for users with higher self-regulation, more subtle interventions, such as the visual gradual intervention, might be preferred.

% The strong negative correlation between impulsivity and self-control in our sample ($r = -.72$; \autoref{sec:construct_overlap}), consistent with prior work~\cite{mao_self-control_2018}, raises the question of whether these constructs are redundant. However, ``\textit{Impulsivity is not merely a lack of self-control, but rather a manifestation of relatively high levels of appetitive motivation for which self-control may serve as a buffer [...]}''~\cite[p. 74]{mao_self-control_2018}. \citet{nigg_annual_2017} further supports this view, distinguishing self-control as the capacity to override impulses from impulsivity as the tendency to act on immediate urges.

\subsubsection{Limited Interaction Between Contextual Factors and Intervention Type}
Among all contextual factors, only sleepiness showed a likely existing interaction effect with intervention type. For sleepy participants, all interventions were similarly effective, whereas alert participants disengaged faster with the baseline pop-up. These findings resonate with research on bedtime procrastination, where people continue late-night scrolling despite intending to stop, often due to reduced self-regulatory capacity~\cite{Kroese.2014}. In such situations, users' intent to continue scrolling may override any intervention regardless of its design. Yet while this effect likely existed, it did not reach significance (2.83\% in ROPE). The limited moderation of the contextual factors aligns with \citet{mongeroffarello2019race}, who argue that ``\textit{users consider their behaviors problematic independently of their contextual situation}'' \cite[p. 11]{mongeroffarello2019race}. Yet, this contrasts prior work showing that contextual factors such as valence and sleepiness do affect overall intervention effectiveness as main effects~\cite{meinhardt_scrolling_2025}. Our results suggest that these two findings are compatible, as contextual factors may influence the overall effectiveness of interventions (main effects), but they do not appear to determine which of our three intervention types is more effective than the others (interaction effects). Hence, stable individual differences appear to be more dominant than momentary contextual factors. 

% This suggests that intervention tailored to participants' self-regulation ability may be a more reliable foundation for digital well-being tools than purely context-based adaptation. Yet, we only measured single moments during one week, missing longer-term effects that may accumulate over weeks or months. 

% \subsection{Interpreting the Limited Impact of Design Friction Interventions in This Study}
\subsection{The Baseline Outperformed the Gradual Interventions}
The baseline pop-up showed higher median values than both gradual interventions across agency and responsiveness (see \autoref{app:descriptive}), diverging from prior design-friction insights~\cite{lu_interactout_2024, ruiz_design_2024}. One explanation for the slower time-to-stop-scrolling might concern insufficient salience of the gradual interventions. As \citet{cox_design_2016} emphasize, interventions must interrupt automatic behavior and trigger deliberate thought. When friction is too subtle, users may adapt without stopping their habituated scrolling. Our results suggest that despite the growing call for subtle design friction in digital well-being~\cite{cox_design_2016, ruiz_design_2024, mejtoft_design_2019, benford_uncomfortable_2012}, participants with low self-control or high impulsivity responded better to the explicit baseline pop-up than to the gradual interventions (see \autoref{fig:lineplots_interaction_effects_objective}). For these users, the baseline pop-up might have provided the clear interruption needed to break infinite scrolling, while the gradual interventions did not. Consequently, moving away from explicit intervention toward more subtle friction designs may actively undermine intervention efficacy for those who need it most.

% A second explanation relates to predictability. \citet{lyngs_self-control_2019} argue that interventions enhance the \textit{expected value of control} when they are predictable and reversible. Our pop-up met both criteria: it appeared in a consistent format and was immediately dismissible. Our gradual interventions, in contrast, were irreversible: once friction increased, the only way to dismiss it was to stop scrolling, likely reducing perceived control and contributing to the lower agency ratings.

% Nevertheless, the primary contribution of this work lies in the interaction between intervention type and contextual or personality factors, rather than the absolute effectiveness of any single intervention.

\subsection{Balancing Subjective and Objective Effectiveness in Intervention Designs}
Our findings reveal a central design dilemma: should designers create interventions for objectively reducing screen time or for a subjective positive experience? Prior work has cautioned against evaluating digital well-being interventions solely by screen time~\cite{Lukoff_online, hiniker2016mytime, Almoallim_2022}. 
Our results also show a divergence between objective and subjective effectiveness: interventions that quickly reduce scrolling are perceived as frustrating, while those rated more positively often delay disengagement. This tension can be interpreted through the distinction between the \textit{experiencing self} and the \textit{remembering self}~\cite{kahneman_thinking_2024}. The experiencing self evaluates interventions in the moment, where abrupt disruptions like pop-ups may feel intrusive or irritating. The remembering self reflects afterwards on whether the intervention ultimately supported meaningful goals, such as spending time with friends~\cite{Terzimehic.2022b}. This suggests that effective intervention must not only nudge users to stop scrolling, but also be framed so that users can look back on the experience as supportive rather than coercive. Designing for the remembering self may justify interventions that momentarily frustrate, as long as they contribute to longer-term well-being.

\subsection{Toward Tailored Interventions}\label{sec:tailored_interentions}
Our findings indicate that individual differences in self-regulation, specifically impulsivity and self-control, moderate the effectiveness of the three interventions we tested. If this moderation pattern generalizes beyond our three designs, it would suggest that interventions could benefit from being adapted to users' self-regulation abilities rather than being deployed uniformly. Related work on adaptive interventions for smartphone overuse has already demonstrated that tailoring intervention timing to contextual factors such as location, activity, and app usage patterns can improve intervention accuracy~\cite{orzikulova_time2stop_2024}. Our findings add to this by indicating that stable individual traits may also inform the selection of intervention type. 
A central challenge for this tailoring is identifying relevant traits without requiring users to complete questionnaires. Recent work suggests that smartphone-usage data itself can serve as a proxy for stable individual differences. For instance, \citet{stachl_predicting_2020} demonstrated that app usage and communication patterns can predict users' Big Five personality traits, particularly conscientiousness and extraversion. Similarly, \citet{wen_mpulse_2021} showed that impulsivity can be inferred from passively collected phone data such as call logs and charging behavior. Because conscientiousness is closely linked to self-control~\cite{nilsen_personality_2024}, passively detected personality traits could provide an indirect predictor for self-regulation ability. This opens up the possibility of adaptive systems that adjust intervention type or intensity to individual needs. For example, offering more explicit interventions to highly impulsive users while relying on subtler gradual friction for those with greater self-control. Such trait-based tailoring would also be efficient to operate. Because self-regulation traits are stable, the inference step needs to run only rarely. This contrasts with context-driven adaptation, where systems must sense the user's context at every moment to decide when or how to intervene~\cite{orzikulova_time2stop_2024}. Thus, continuous sensing incurs recurring costs in battery power, computation, and access to potentially sensitive signals such as location or activity. Our results suggest that, at least for selecting the intervention type, this cost may be avoidable, as the contextual factors we measured showed little moderating influence, whereas stable traits did. 

% Evidence from other domains of behavior change shows that gradual and adaptive approaches can be highly effective. For example, smoking cessation programs often use gradual nicotine reduction~\cite{lindson_smoking_2019}.
% , while studies on physical activity highlight that individual traits influence how well people respond to different interventions~\cite{stieger_relationship_2020}. 
% In the mental health domain, outcomes also vary by individual differences: conscientious users benefit more from structured activities such as journaling, whereas less conscientious individuals respond better to playful interventions like games~\cite{khwaja_personality_2021}. These parallels highlight that digital well-being interventions could benefit from adopting principles of gradual intensity increase and tailoring, which are already used in health domains. 

\subsection{Limitations}
Several limitations should be considered. First, our participant pool was limited to Android users in the US and UK, excluding iOS users, which limits the generalizability of the results. Second, our re-engagement analysis (see \autoref{sec:reengagement}) showed that 13.7\% of consecutive sessions resumed within 5\,min. Such rapid returns may reflect natural scrolling behavior (e.g., briefly replying to a message before continuing to scroll), but could also indicate attempts to intentionally increase compensation by triggering additional data points. Because rapid re-engagement may also indicate limited intervention effectiveness, we retained all such data to avoid artificially removing behaviors that may occur in real use. However, this choice prevents us from distinguishing re-engagement from incentive-driven behavior, limiting the precision of our behavioral interpretation. A related measurement limitation concerns our use of time-to-stop as objective effectiveness. This measure inherently reflects both the time needed to perceive an intervention and the time needed to decide to stop scrolling. Since gradual interventions introduce friction gradually, their effects may unfold over a longer perceptual timescale than sudden pop-ups. As a result, direct comparisons across these intervention types are imperfect, and responsiveness may systematically advantage abrupt cues. We nevertheless used this metric because it is one of the standard metrics in intervention research~\cite{orzikulova_time2stop_2024, hiniker2016mytime, kim2019lockntype, lu_interactout_2024, meinhardt_scrolling_2025}. 
An additional limitation concerns our ESM design. Because the subjective effectiveness questionnaire was triggered only after participants stopped scrolling, all subjective ratings and contextual reports are conditioned on the decision to disengage. This introduces a potential self-selection bias, as differences in subjective ratings across interventions or context may partly reflect who stopped rather than how effective the intervention was perceived to be. Nonetheless, we chose this post-session timing because collecting subjective measures during scrolling would have interrupted the activity under study.

Our study examined only three interventions grounded in prior work~\cite{kincaid_gradual_2023, okeke2018good, ruiz_design_2024}, representing a narrow subset of possible friction designs. Their specific parameters (e.g., the 3\,min 31\,s linear increase, visual/haptic designs, or the 15~min trigger threshold) represent only one design in a larger space. Alternative friction designs, such as faster/slower increases, different slopes, or hybrid designs combining pop-ups with gradual friction, might yield different results. Thus, our findings highlight the importance of specific design choices rather than implying general implications of friction-based approaches. Still, the main effects of individual differences that emerged independently of the intervention type (see \autoref{app:bayesian_model}) suggest that, besides our limited design space, individual differences shape how users respond to interventions. Further, because intervention types were assigned by full randomization at each decision point, 46 participants received three or more interventions in total yet were never exposed to all three types. A counterbalanced randomization system that cycles through all conditions before repeating any type would have prevented this attrition while still preserving the benefits of random assignment.

Finally, the study lasted only 7 days. While this ensured ecologically valid data collection, longer-term effects such as habituation remain unexplored~\cite{kovacs_rotating_2018}, and meaningful behavioral changes typically emerge over several weeks (e.g., 13 weeks~\cite{haliburton_longitudinal_2024}). The short study duration may also blur the comparison of intervention types, because the designs differed in how familiar they were to participants. The baseline is based on the screen-time reminders that platforms already show their users~\cite{TikTokScreenTime2025, InstagramTimeLimit2025}, whereas the gradual interventions were new to participants. Part of the baseline's advantage in responsiveness and ratings may therefore stem from users recognizing a familiar intervention rather than from the design itself, an advantage that could fade as users grow equally accustomed to the gradual interventions. Over longer durations, these differences could shift. The trait moderation, in contrast, rests on stable individual differences and should be less sensitive to the study duration, yet whether it persists over months remains open. Further, habituation poses a particular challenge for non-adaptive interventions, as users may initially respond to them but gradually ignore them as novelty fades~\cite{kovacs_rotating_2018}. This limitation strengthens the case for interventions that can dynamically adjust as users become habituated, potentially maintaining effectiveness over longer-term use.

\subsection{Future Work}
Future work should explore a broader design space of intervention designs. An interesting approach to optimize these design parameters could be Multi-Objective Bayesian Optimization, an algorithmic method that efficiently explores design parameter spaces to optimize for objective and subjective intervention effectiveness. This method showed success in related UI optimization challenges~\cite{brochu2010bayesian, kadner2021adaptifont, dudley2019crowdsourcing}. 
Whether our moderation findings translate into practical benefits when used to tailor interventions to individual traits remains an open question. Since such traits can be inferred from phone usage patterns~\cite{wen_mpulse_2021, stachl_predicting_2020}, future work should investigate whether adaptive systems that dynamically adjust intervention parameters based on detected traits outperform fixed intervention designs. However, the same technology that enables personalization also enables surveillance, requiring careful attention to user privacy. Therefore, adaptive interventions should strike a balance between tailoring and privacy, prioritizing people's well-being over optimizing for (dis-)engagement alone.

\subsection{Policy Implications}
Our findings have implications for ongoing regulatory debates, especially those related to Article 28 of the Digital Services Act~\cite{DSA}, which requires online platforms to ensure a high level of privacy, safety, and security for minors. The accompanying guidelines specify what this means for intervention design. They recommend ``\textit{information or friction that slows down content display [...] giving users an opportunity to think before they decide if they want to see more content}''~\cite[§57.b.ix]{DSA}, and at the same time define interventions as effective if they``\textit{deter minors from spending more time on the platform}''~\cite[§61.c]{DSA}. Therefore, the guidelines define intervention effectiveness as reduced usage time. Our results suggest this criterion is too narrow. Participants with low self-control and high impulsivity benefited from the explicit baseline pop-up but not from either gradual intervention (see
\autoref{sec:baysian_model}), even though gradual designs are the kind of low-friction implementation a usage-reduction criterion would most readily accept. Additionally, since users dismiss or abandon unacceptable interventions~\cite{Okeke.2018, lukoff2022designing}, platforms have their own incentive to favor seamless designs. A usage-reduction criterion does not counter this, leaving limited pressure to develop stronger designs for the users most at risk.

We therefore suggest two refinements. First, audits should report outcomes separately for user groups defined by self-regulation traits such as self-control and impulsivity, and should combine objective behavior change with subjective measures, rather than relying on population averages or on usage reduction alone. Second, regulation should specify measurable outcomes, such as low rates of intervention dismissal, rather than prescribing particular designs, since no single design in our study served every user equally.

\section{Conclusion}
This paper investigated whether the effectiveness of different intervention types for mitigating infinite scrolling depends on users' context and individual traits. In a 7-day field study with $N=104$ participants, we compared a baseline pop-up with gradually intensifying visual and haptic interventions. By investigating both objective and subjective intervention effectiveness, we found that users responded differently to the three interventions: the baseline pop-up yielded the fastest disengagement but quickly lost influence if ignored, the haptic intervention disrupted scrolling earlier but was subjectively perceived as less effective as intensity increased, and the visual intervention was slower to trigger disengagement but maintained higher subjective effectiveness over time. These results highlight that interventions that effectively reduce screen time may not always be what users find acceptable. Our Bayesian analysis further showed that individual differences in self-regulation, specifically self-control and impulsivity, shaped how participants responded to the interventions, while contextual factors showed only limited moderation of which intervention worked most effectively. Participants with low self-control benefited most from the baseline pop-up, which offered an explicit prompt to stop scrolling, while those with higher self-control were less dependent on our tested intervention types. 
% This suggests providing explicit interventions for users with low self-regulation while relying on gradual, less disruptive friction for those with greater self-control. 
Overall, our findings show that intervention design should balance objective impact with subjective experience and account for differences in individuals' self-regulation traits. However, because both gradual interventions were outperformed by a simple pop-up in our study, the moderation effects we observed should be understood as motivation for future research on tailored interventions rather than as direct evidence that tailoring improves outcomes.

\section*{Open Science}
The Android application used for this study, the study data, and the R scripts used for analysis are openly available at:
\url{https://github.com/luca-maxim/Cant_Stop}.
To protect participant privacy, all study data is shared in anonymized form, with Prolific IDs replaced by unique sequential participant identifiers.

\begin{acks}
This research was supported by the German Research Foundation (DFG) through the project "Beyond Screen Time: Context- and Content-tailored Interventions to Social Media Usage to Enhance Digital Well-being " (project number: \href{https://gepris.dfg.de/project/561828495?lang=en}{561828495}). We are further grateful to Lukas Gruler for building an initial web-based feasibility study of the gradual interventions, to Maryam Elhaidary for drafting early Android implementations of the gradual interventions, and to Albin Zeqiri for exploring first approaches to analyzing the user data. Thanks to RHCP for the main theme of this work.
\end{acks}

%TC:ignore

\bibliographystyle{ACM-Reference-Format}
\bibliography{sample-base}

% \newpage
\appendix

\section{Descriptive Data of the User Study}\label{app:descriptive}

\begin{figure}[ht]
    \centering
    \includegraphics[width=0.56\linewidth]{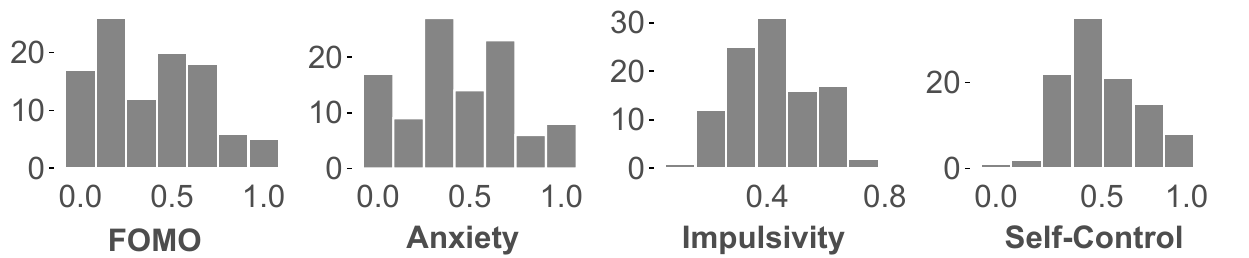}
    \caption{Distribution of the between-factors variables of the participants (normalized)}~\label{fig:between_distribution}
   \Description{These plots show the distributions of the contextual factors assessed during the user study}
    \label{fig:between_distribution}
\end{figure}
\vspace{-0.6cm}
\begin{figure}[ht]
    \centering
    \includegraphics[width=0.76\linewidth]{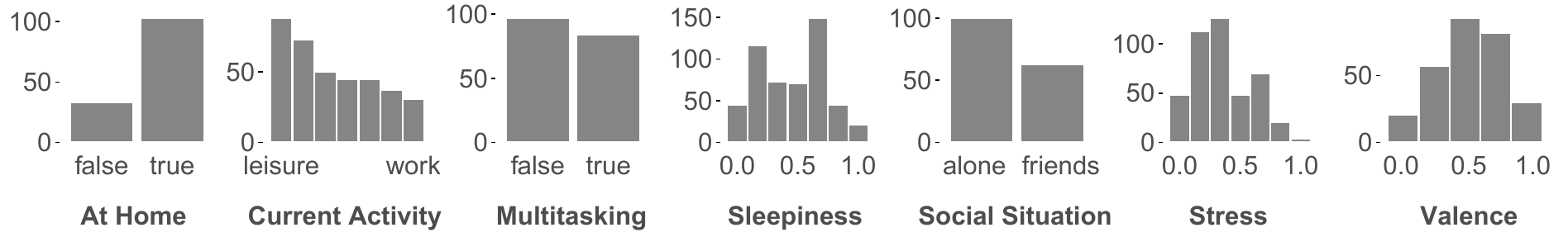}
    \caption{Distribution of the within-factors variables of the participants (normalized)}~\label{fig:within_distribution}
   \Description{These plots show the distributions of the between-factors assessed prior to the user study}
    \label{fig:within_distribution}
\end{figure}
\vspace{-0.6cm}
\begin{figure}[ht]
    \centering
    \includegraphics[width=0.66\linewidth]{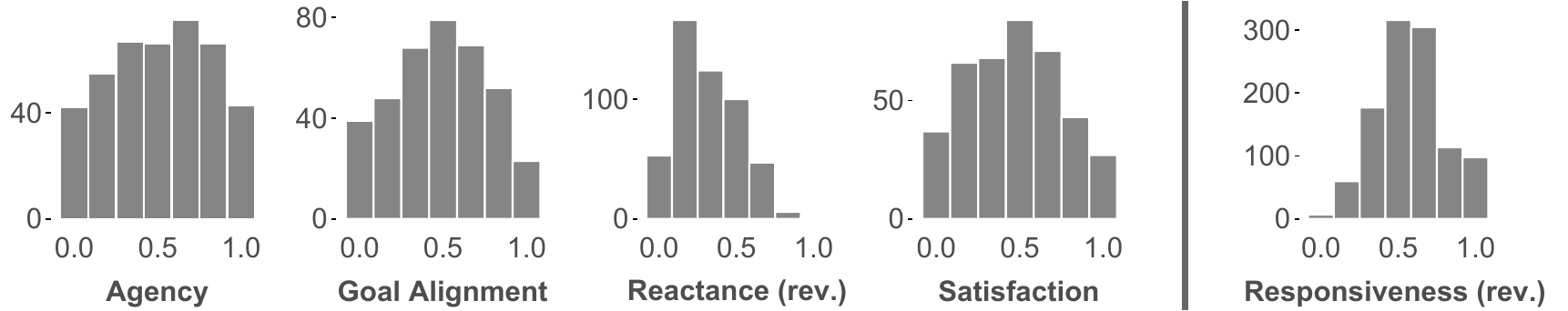}
    \caption{Distribution of the six measures (normalized) that were used to assess objective and subjective intervention effectiveness. reactance and responsiveness are reversed (see \autoref{app:usedquestionaires})}~\label{fig:objective_distribution}
   \Description{These plots show the distributions of the subjective and objective measures to assess intervention effectiveness}
\end{figure}

\clearpage

\begin{table}[ht!]
\small
\caption{Descriptive statistics for contextual factors, individual traits, and intervention effectiveness. All values are normalized, except for responsiveness.}
\begin{tabular}{lrrrrrl}

 \textbf{Contextual Factor} & \textbf{Min} & \textbf{Max} & \textbf{Mean} & \textbf{Median} & \textbf{SD} & \textbf{Distribution} \\ 
    \hline 
   \addlinespace[1pt]
   \hline 
  \addlinespace[2pt]
    Sleepiness & 0.00 & 1.00 & 0.46 & 0.50 & 0.28 & \\
    Current Activity & 0.00 & 1.00 & 0.27 & 0.17 & 0.31 & \\
    Valence & 0.00 & 1.00 & 0.55 & 0.50 & 0.23 & \\
    At Home &  &  &  & & & true (89.72\%), false (10.28\%)\\
    Multitasking &  &  & &  &  & true (40.42\%), false (59.58\%) \\
    Social Situation &  &  &  &  &  & alone (72.80\%), friends (27.20\%) \\
    \\
     \addlinespace[-7pt]
    \textbf{Individual Traits} &  & &  &  &  &  \\ 
  \hline
  \addlinespace[2pt]
    Self-Control & 0.06 & 0.92 & 0.50 & 0.48 & 0.19 & \\
    Anxiety & 0.00 & 1.00 & 0.43 & 0.47 & 0.29 & \\
    FOMO & 0.00 & 1.00 & 0.41 & 0.35 & 0.28 & \\
    Impulsivity & 0.13 & 0.76 & 0.41 & 0.40 & 0.14 & \\
    \\
    \addlinespace[-7pt]
   \textbf{Intervention Effectiveness} &  &  &  &  &  &  \\
   \hline 
   \addlinespace[2pt]
   \multicolumn{7}{l}{\textit{Subjective Effectiveness}} \\
   \quad Baseline & 0.00 & 0.98 & 0.48 & 0.51 & 0.24 & \\
   \quad Visual Intervention & 0.00 & 0.92 & 0.40 & 0.42 & 0.22 & \\
   \quad Haptic Intervention & 0.00 & 0.97 & 0.43 & 0.43 & 0.25 & \\
   \addlinespace[2pt]
   \multicolumn{7}{l}{\textit{Objective Effectiveness}} \\
   \quad Baseline & 0.00 & 1.00 & 0.65 & 0.71 & 0.30 & \\
   \quad Visual Intervention & 0.03 & 1.00 & 0.51 & 0.44 & 0.26 & \\
   \quad Haptic Intervention & 0.00 & 1.00 & 0.55 & 0.54 & 0.25 & \\
   \addlinespace[2pt]
   \multicolumn{7}{l}{\textit{Individual measures}} \\
   \quad Responsiveness -- Baseline & 0s & 23min 10s & 2min 2s & 7s & 4min 40s & \\
   \quad Responsiveness -- Visual & 0s & 18min 10s & 1min 42s & 56s & 2min 28s & \\
   \quad Responsiveness -- Haptic & 0s & 23min 13s & 1min 31s & 28s & 2min 52s & \\
   \addlinespace[2pt]
   \quad Reactance (rev.) -- Baseline & 0.10 & 1.00 & 0.72 & 0.75 & 0.20 & \\
   \quad Reactance (rev.) -- Visual & 0.00 & 1.00 & 0.74 & 0.75 & 0.19 & \\
   \quad Reactance (rev.) -- Haptic & 0.20 & 1.00 & 0.76 & 0.75 & 0.20 & \\
   \addlinespace[2pt]
   \quad Goal Alignment -- Baseline & 0.00 & 1.00 & 0.51 & 0.50 & 0.29 & \\
   \quad Goal Alignment -- Visual & 0.00 & 1.00 & 0.43 & 0.50 & 0.27 & \\
   \quad Goal Alignment -- Haptic & 0.00 & 1.00 & 0.47 & 0.50 & 0.30 & \\
   \addlinespace[2pt]
   \quad Satisfaction -- Baseline & 0.00 & 1.00 & 0.51 & 0.50 & 0.29 & \\
   \quad Satisfaction -- Visual & 0.00 & 1.00 & 0.40 & 0.42 & 0.28 & \\
   \quad Satisfaction -- Haptic & 0.00 & 1.00 & 0.44 & 0.50 & 0.31 & \\
   \addlinespace[2pt]
   \quad Agency -- Baseline & 0.00 & 1.00 & 0.59 & 0.67 & 0.31 & \\
   \quad Agency -- Visual & 0.00 & 1.00 & 0.45 & 0.50 & 0.31 & \\
   \quad Agency -- Haptic & 0.00 & 1.00 & 0.53 & 0.50 & 0.33 & \\
   \addlinespace[2pt]
   \quad Usefulness -- Baseline & 0.00 & 1.00 & 0.49 & 0.50 & 0.33 & \\
   \quad Usefulness -- Visual & 0.00 & 1.00 & 0.41 & 0.33 & 0.30 & \\
   \quad Usefulness -- Haptic & 0.00 & 1.00 & 0.44 & 0.50 & 0.33 & \\
   \\
   \addlinespace[-7pt]
   \textbf{Intervention Distribution} &  &  & &  &  &  \\
   \hline
   \addlinespace[2pt]
   \multicolumn{7}{l}{Baseline (34.23\%), Visual Intervention (33.69\%), Haptic Intervention (32.07\%)} \\
   \\
   \addlinespace[-7pt]
   \textbf{App Distribution} &  &  & &  &  &  \\
   \hline
   \addlinespace[2pt]
   \multicolumn{7}{l}{TikTok (41.27\%), Instagram (22.91\%), Facebook (17.91\%), YouTube Shorts (17.91\%)} \\
   
\label{tab:_descr_stat}
\end{tabular}
\end{table}

\clearpage

\begin{figure}[ht!]
    \centering
    \includegraphics[width=0.8\linewidth]{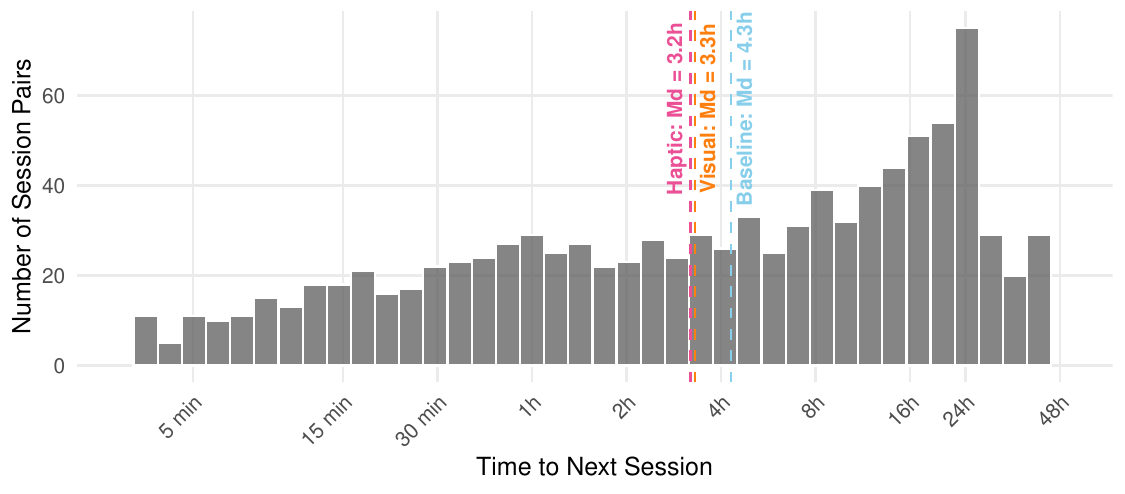}
    \caption{Distribution of break times between consecutive (intervention-triggering) scrolling sessions. Each dotted line represents the median break time per intervention type before the next session. Only sessions exceeding the 15-minute threshold (triggering an intervention) are included.}~\label{fig:reengagement_histogram}
   \Description{This histogram shows the distribution of break times between consecutive intervention-triggered scrolling sessions. Bars span from 5 minutes to 48 hours. Three vertical dashed lines mark the median time to next session for each intervention: visual (Md=3.3h), haptic (Md=3.2h), and baseline (Md=4.3h).}
\end{figure}

\section{Question Items Used in the User Study}\label{app:usedquestionaires}

\begin{table*}[ht]
\small 
\caption{Question items used in the user study to determine \textbf{context}}
\begin{tabularx}{\textwidth}{p{2.5cm}p{6cm}p{5.3cm}l}

   \textbf{Measurement}& Question Item & Answer Items & Ref.   \\
   \hline 
   \addlinespace[0.6pt]

   \hline 
   \addlinespace[1.5pt]
   \textbf{Current Activity}
   & What is your current activity?
   & 7-point Likert-type scale from -3 (“definitely leisure”) to 3 (“definitely not leisure”) & \cite{Samdahl.1991} \\

    \addlinespace[0.6pt]
   \hline 
   \addlinespace[0.6pt]
   \textbf{Valence}
   & How do you feel?
   & five images of manikin showing different valence levels & \cite{Bradley.1994}   \\
   
   \addlinespace[0.6pt]
   \hline 
   \addlinespace[0.6pt]
   \textbf{Sleepiness}
   & What is your level of sleepiness?
   & 9-point Likert-type scale from 1 (“extremely alert”) to 9 (“extremely sleepy”) & \cite{Shahid.2012b} \\

\addlinespace[0.6pt]
   \hline 
   \addlinespace[0.6pt]
   \textbf{Social Situation}
   & Which one of these best describes people around you?
   & “alone”, “with friends/colleagues/family members” & \cite{Akpinar.2023} \\

   \addlinespace[0.6pt]
   \hline 
   \addlinespace[0.6pt]
   \textbf{Stress}
   & What number best describes your level of stress right now?
   & 11-point Likert-type scale, ranging from 0 (“no stress”) to 10 (“worst stress possible”) & \cite{Karvounides2016} \\

   \addlinespace[0.6pt]
   \hline 
   \addlinespace[0.6pt]
   \textbf{Multitasking}
   & Did you do anything else besides being on [app name]?
   & “yes”, “no” & \cite{meinhardt_scrolling_2025} \\

   \addlinespace[0.6pt]
   \hline 
   \addlinespace[0.6pt]
   \textbf{At Home}
   & Are you currently at home?
   & “yes”, “no” & \cite{meinhardt_scrolling_2025} \\

\label{tab:question_items}
\end{tabularx}
\end{table*}

\begin{table*}[ht!]
\small 
\caption{Measures and questionnaire items to determine subjective and objective \textbf{intervention effectiveness}}
\begin{tabularx}{\textwidth}{p{2.4cm}p{6.4cm}p{2.6cm}lX}

   \textbf{Measurement}& Question Item & Answer Items & Ref. & Calculation   \\
   \hline 
\addlinespace[0.7pt]
   \hline 
   \addlinespace[3.4pt]
   \multicolumn{5}{l}{\textbf{Objective Effectiveness}} \\
   \hline 
   \addlinespace[0.7pt]
   \hline
   \addlinespace[1pt]
   \quad Responsiveness
   & 
   & & \cite{meinhardt_scrolling_2025}
   & $log( 1 + Responsiveness)$, \newline normalized (0-1), and reversed\\

      \addlinespace[-8pt]
   \multicolumn{5}{l}{\textbf{Subjective Effectiveness} (weighted measures)} \\
   \hline 
   \addlinespace[0.7pt]
   \hline
   \addlinespace[1pt]

   \addlinespace[1pt]
   \quad Reactance
   & 
   I want to be in control, not the intervention. \newline
   I like to act independently from the intervention. \newline
   I don't want the intervention to tell me what to do. \newline
   I don't let the intervention impose its will on me. \newline
   I alone determine what to do, not the intervention.
   &    5-point Likert scale from\newline
 “strongly disagree”, to “strongly agree” & \cite{Ehrenbrink.2020} 
 & normalized (0-1) and reversed \\
   
   \addlinespace[0.7pt]
   \hline 
   \addlinespace[1pt]
   \quad Agency
   & For this intervention, how much did you feel out of or in control?
   & 7-point Likert-type scale, ranging from 1 (“very out of control”)  to 7 (“very in control”) & \cite{kai_internal}
   & normalized (0-1)\\

\addlinespace[0.7pt]
   \hline 
   \addlinespace[1pt]
   \quad Satisfaction
   & For this intervention, how much did you feel dissatisfied or satisfied?
   & 7-point Likert-type scale, ranging from 1 (“very dissatisfied”)  to 7 (“very satisfied”) & \cite{kai_internal} 
   & normalized (0-1) \\

   \addlinespace[0.7pt]
   \hline 
   \addlinespace[1pt]
   \quad Goal Alignment
   & For this intervention, how much did it conflict with or support your personal goals?
   & 7-point Likert-type scale, ranging from 1 (“very in conflict”) to 7 (“very supported”) & \cite{kai_internal} 
   & normalized (0-1)\\

    \addlinespace[0.7pt]
   \hline 
   \addlinespace[1pt]
   \quad Usefulness
   & I find the intervention to be useful in my current situation.
   & 7-point Likert-type scale, 
ranging from 1 (“strongly disagree”)  
to 7 (“strongly agree”) & \cite{Venkatesh2000} 
& normalized (0-1) \\ 

\label{tab:question_items}
\end{tabularx}
\end{table*}

\begin{table*}[h]
\small 
\caption{Question items used in the user study to determine \textbf{individual traits}}
\begin{tabularx}{\textwidth}{p{1.9cm}p{9cm}p{2.8cm}l}

   \textbf{Measurement}& Question Item & Answer Items & Ref.   \\
   \hline 
   \addlinespace[1pt]
   \hline 
   \addlinespace[1pt]
   \textbf{Impulsivity}
   & I act on impulse. [inverted]\newline  % inverted ???
   I act on the spur of the moment.\newline
    I do things without thinking.\newline
    I say things without thinking.\newline
    I buy things on impulse.\newline
    I plan for job security. [inverted]\newline
    I plan for the future. [inverted]\newline
    I save regularly. [inverted]\newline
    I plan tasks carefully. [inverted]\newline
    I am a careful thinker. [inverted]\newline
    I am restless at lectures or talks.\newline
    I squirm at plays or lectures.\newline
    I concentrate easily. [inverted]\newline
    I don't pay attention.\newline
    Easily bored solving thought problems.
   &4-point Likert scale, 
ranging from 1 (“rarely/never”) 
to 4 (“almost always”) & \cite{Spinella2007} \\

    \addlinespace[1pt]
   \hline 
   \addlinespace[1pt]
   \textbf{Anxiety}
   & I feel that difficulties are piling up so that I can't overcome them.\newline
I worry too much over something that really doesn’t matter.\newline
Some unimportant thought runs through my mind and bothers me.\newline
I take disappointments so keenly that I can’t put them out of my mind.\newline
I get in a state of tension or turmoil as I think over my recent concerns and interests.
   & 4-point Likert scale, ranging from 0 (“almost never”) to 3 (“almost always”) & \cite{Zsido2020}   \\
   
   \addlinespace[1pt]
   \hline 
   \addlinespace[1pt]
   \textbf{FOMO}
   & I fear others have more rewarding experiences than me.\newline
I fear my friends have more rewarding experiences than me.\newline
I get worried when I find out my friends are having fun without me.\newline
I get anxious when I don’t know what my friends are up to.\newline
When I miss out on a planned get-together it bothers me.
   & 5-point Likert scale, ranging from 1 (“totally disagree”) to 5 (“totally agree”) & \cite{Wegmann2017} \\
   
   \addlinespace[1pt]
   \hline 
   \addlinespace[1pt]
   \textbf{Self-Control}
   & I am good at resisting temptation.\newline
I have a hard time breaking bad habits. [inverted]\newline
I am lazy. [inverted]\newline
I say inappropriate things. [inverted]\newline
I do certain things that are bad for me, if they are fun. [inverted]\newline
I refuse things that are bad for me.\newline
I wish I had more self-discipline. [inverted]\newline
People would say that I have iron self-discipline.\newline
Pleasure and fun sometimes keep me from getting work done. [inverted]\newline
I have trouble concentrating. [inverted]\newline
I am able to work effectively toward long-term goals.\newline
Sometimes I can't stop myself from doing something, even if I know it is wrong. [inverted]\newline
I often act without thinking through all the alternatives. [inverted]
   & 5-point Likert scale,
ranging from 1 (“not at all”)
to 5 (“very much”)
   & \cite{tangney_high_2004} \\
\label{tab:question_items}
\end{tabularx}
\end{table*}

\clearpage

\section{Mixed Power Simulation}\label{app:power_sim}
To determine the required sample size for sufficient statistical power, we conducted a simulation-based power analysis using pilot data from the first batch of 30 participants, following the approach by \citet{Kumle.2021}. However, simulating the full model with all contextual variables was computationally intensive. To address this, we first fitted the full model on the data of the 30 participants and identified factors with $|t| > 2$ (see \autoref{tab:xgb-feature-importance}a). 
We then limited the power simulation to those factors where $|t|>2$~\cite{Kumle.2021}: anxiety, impulsivity, and FOMO. The simulation indicated that a sample size of 90 participants would be sufficient to achieve 80\% power~\cite{cohen_power_1992} (see \autoref{tab:xgb-feature-importance}b).

\begin{table}[H]
\small
\centering
\caption{Feature importance (a) and corresponding power estimates (b) for included predictors and their interactions}
\label{tab:feature-power-side-by-side}
\begin{minipage}[t]{0.4\textwidth}
\centering
\caption*{\textbf{(a) t-values of the full model's interactions on the pilot data (N = 30)}}
\label{tab:xgb-feature-importance}
\begin{tabularx}{\textwidth}{lX}
\toprule
\textbf{Interaction Effect} & \textbf{t-value} \\
\midrule
Visual Inter. × Valence & -0.32 \\
Haptic Inter. × Valence & 0.11 \\
\addlinespace[3pt]
\textbf{Visual Inter. × Impulsivity} & \textbf{-2.69} \\
Haptic Inter. × Impulsivity & 0.22 \\
\addlinespace[3pt]
Visual Inter. × Sleepiness & -0.70 \\
Haptic Inter. × Sleepiness & -0.59 \\
\addlinespace[3pt]
Visual Inter. × FOMO & 1.20 \\
\textbf{Haptic Inter. × FOMO} & \textbf{2.22} \\
\addlinespace[3pt]
Visual Inter. × Stress & 0.75 \\
Haptic Inter. × Stress & -1.21 \\
\addlinespace[3pt]
Visual Inter. × Multitasking & -0.37 \\
Haptic Inter. × Multitasking & 0.65 \\
\addlinespace[3pt]
Visual Inter. × At Home & 0.05 \\
Haptic Inter. × At Home & -1.19 \\
\addlinespace[3pt]
Visual Inter. × Self-Control & 0.47 \\
Haptic Inter. × Self-Control & -0.02 \\
\addlinespace[3pt]
\textbf{Visual Inter. × Anxiety} & \textbf{3.51} \\
Haptic Inter. × Anxiety & -1.15 \\
\addlinespace[3pt]
Visual Inter. × Current Activity & -0.88 \\
Haptic Inter. × Current Activity & -0.63 \\
\addlinespace[3pt]
Visual Inter. × Social Situation & -1.63 \\
Haptic Inter. × Social Situation & -0.08 \\
\bottomrule
\end{tabularx}
\end{minipage}
\hfill
\begin{minipage}[t]{0.55\textwidth}
\centering
\caption*{\textbf{(b) Power Simulation Estimates for Interactions (based on 1000 simulation runs)}}
\begin{tabularx}{\textwidth}{lcc:ccc}
\toprule
\textbf{Interaction Effect} & \multicolumn{5}{c}{\textbf{Power per Sample Size}} \\
\cmidrule(lr){2-6}
               & 30   & 60   & 90   & 120  & 150  \\
\midrule
Visual Interv. × Anxiety     & 0.43 & \textbf{0.92} & \textbf{0.99} & \textbf{0.99} & \textbf{1.00} \\
Haptic Interv. × Anxiety     & 0.12 & 0.31 & 0.45 & 0.55 & 0.67 \\
\hline
Visual Interv. × Impulsivity & \textbf{0.92} & \textbf{1.00} & \textbf{1.00} & \textbf{1.00} & \textbf{1.00} \\
Haptic Interv. × Impulsivity & 0.04 & 0.06 & 0.05 & 0.04 & 0.06 \\
\hline
Visual Interv. × FOMO        & 0.31 & 0.74 & \textbf{0.89} & \textbf{0.96} & \textbf{0.99} \\
Haptic Interv. × FOMO        & 0.28 & 0.73 & \textbf{0.88} & \textbf{0.95} & \textbf{0.98} \\
\bottomrule
\end{tabularx}

\end{minipage}
\end{table}

\clearpage

\section{Results of the Bayesian Model}\label{app:bayesian_model}

\begin{figure}[ht!]
\centering
\small
    \begin{subfigure}[c]{0.5\linewidth}
        \includegraphics[width=\linewidth]{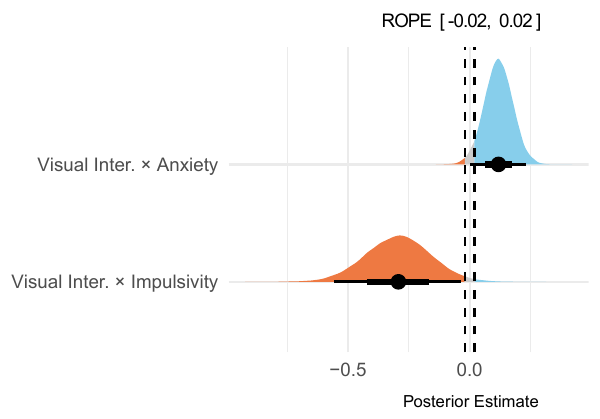}      \caption{Interaction effects on \textbf{subjective effectiveness}}\label{fig:effects_subjectiev_effectiveness}
        \Description{This subfigure displays the subjective effectiveness. The visual intervention interacting with anxiety and impulsivity shows posterior estimates, with the impulsivity effect being negative.}
    \end{subfigure}
    
    \vspace{0.8cm}
    
    \begin{subfigure}[c]{0.5\linewidth}
         \includegraphics[width=\linewidth]{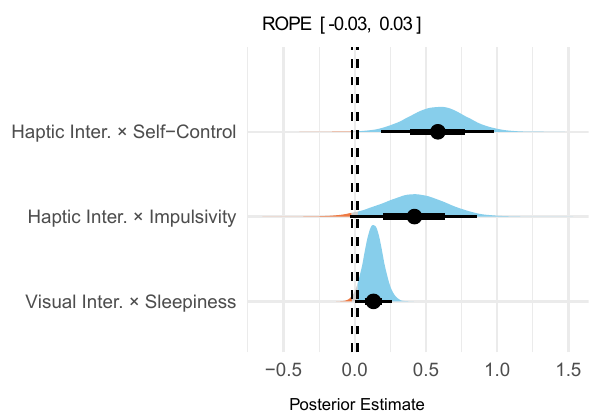}      \caption{Interaction effects on \textbf{objective effectiveness}}\label{fig:effects_objective_effectiveness}
         \Description{This subfigure shows the objective effectiveness. The haptic intervention interacting with self-control and impulsivity shows positive posterior estimates, as does the visual intervention interacting with sleepiness. }
    \end{subfigure} 
   \caption{Interaction Effects between the intervention type and the individual traits where the $pd > 95\%$}
   \label{fig:interations_figure_appx}
   \Description{The figure shows half-eyed plots of interaction effects between intervention type and individual traits on both subjective and objective effectiveness. displays subjective effectiveness. The visual intervention interacting with anxiety and impulsivity shows posterior estimates, with the impulsivity effect being negative. The Region of Practical Equivalence (ROPE) is indicated by vertical dashed lines around zero, with subjective effectiveness using the range -0.02 to 0.02 and objective effectiveness using -0.03 to 0.03. Black dots and intervals show posterior means and uncertainty.}
\end{figure}

We estimated two Bayesian mixed-effects models to examine how intervention type and individual/contextual factors affect objective and subjective intervention effectiveness. This section presents the complete model results in \autoref{tab:results_objective} for the objective effectiveness and \autoref{tab:results_subjective} for the subjective effectiveness of the interventions. \autoref{fig:interations_figure_appx} shows the existing interaction effects (pd > 95\%) between intervention type and individual traits on both subjective and objective intervention effectiveness. The vertical dashed lines indicate the ROPE where effects are considered practically negligible. When the percentage in ROPE is below 2.5\% an effect can be considered as probably significant~\cite{makowski_indices_2019}.

\begin{table*}[ht]
\small
\caption{Results of the Bayesian mixed-effects model for \textbf{subjective effectiveness}. Effects with a \textit{pd} $>$ 95\% are highlighted in bold.}
\label{tab:results_subjective}
\begin{tabular}{lccccc}
\toprule
\toprule
\multicolumn{6}{c}{\textbf{Subjective Effectiveness}} \\
\toprule
\toprule
\addlinespace[4pt]
Parameter & Median & 95\% CrI & pd & \% in ROPE & ESS \\
\midrule
\textit{Intercept} & \textit{0.12} & \textit{[-0.16, 0.40]} & \textit{81.18\%} & \textit{9.56\%} & \textit{18,506} \\
\addlinespace[4pt]
Visual Inter. & -0.06 & [-0.29, 0.16] & 71.12\% & 14.91\% & 18,619 \\
Haptic Inter. & -0.03 & [-0.24, 0.18] & 61.38\% & 18.08\% & 19,4831 \\
\addlinespace[4pt]
\textbf{Valence} & \textbf{0.38} & \textbf{[0.29, 0.46]} & \textbf{100\%} & \textbf{0\%} & \textbf{21,333} \\
\textbf{Sleepiness} & \textbf{0.08} & \textbf{[0.03, 0.13]} & \textbf{99.95\%} & \textbf{0\%} & \textbf{28,091} \\
Stress & 0.02 & [-0.06, 0.10] & 67.97\% & 42.43\% & 26,071 \\
Multitasking [Yes] & 0.00 & [-0.03, 0.03] & 52.98\% & 95.78\% & 26420 \\
At Home [True] & 0.00 & [-0.06, 0.04] & 62.44\% & 63.86\% & 28,445 \\
Current Activity & 0.01 & [-0.04, 0.06] & 69.70\% & 62.03\% & 26,789 \\
Social Situation [Friends] & 0.00 & [-0.04, 0.03] & 52.52\% & 88.33\% & 23,941 \\
\addlinespace[4pt]
\textbf{Impulsivity} & \textbf{0.33} & \textbf{[0.01, 0.67]} & \textbf{97.77\%} & \textbf{0.65\%} & \textbf{15,797} \\
FOMO & -0.11 & [-0.27, 0.05] & 92.03\% & 9.41\% & 24,501 \\
Self-Control & 0.00 & [-0.28, 0.28] & 50.25\% & 14.51\% & 23,342 \\
Anxiety & 0.06 & [-0.08, 0.21] & 80.57\% & 19.19\% & 27,244 \\
\addlinespace[2pt]
\hline
\addlinespace[2pt]
Visual Inter. × Valence & 0.04 & [-0.05, 0.13] & 82.52\% & 28.08\% & 25,621 \\
Haptic Inter. × Valence & 0.02 & [-0.07, 0.11] & 68.30\% & 37.55\% & 25,146 \\
\addlinespace[4pt]
Visual Inter. × Sleepiness & -0.03 & [-0.10, 0.04] & 81.62\% & 38.95\% & 27,279 \\
Haptic Inter. × Sleepiness & 0.00 & [-0.05, 0.07] & 61.76\% & 54.44\% & 31,846 \\
\addlinespace[4pt]
Visual Inter. × Stress & -0.02 & [-0.12, 0.08] & 67.34\% & 35.44\% & 23,862 \\
Haptic Inter. × Stress & -0.04 & [-0.13, 0.05] & 82.66\% & 28.05\% & 29,071 \\
\addlinespace[4pt]
Visual Inter. × Multitasking [Yes] & 0.02 & [-0.02, 0.05] & 80.69\% & 66.75\% & 29,839 \\
Haptic Inter. × Multitasking [Yes] & 0.00 & [-0.03, 0.04] & 63.69\% & 79.53\% & 33,202 \\
\addlinespace[4pt]
Visual Inter. × At Home [True] & 0.01 & [-0.06, 0.07] & 62.09\% & 54.61\% & 30,071 \\
Haptic Inter. × At Home [True] & 0.02 & [-0.04, 0.08] & 74.75\% & 49.97\% & 33,719 \\
\addlinespace[4pt]
Visual Inter. × Current Activity & 0.02 & [-0.04, 0.08] & 72.12\% & 49.22\% & 27,196 \\
Haptic Inter. × Current Activity & 0.00 & [-0.07, 0.06] & 58.94\% & 55.90\% & 27,197 \\
\addlinespace[4pt]
Visual Inter. × Social Situation [Friends] & -0.01 & [-0.05, 0.03] & 73.39\% & 68.89\% & 26,516 \\
Haptic Inter. × Social Situation [Friends] & 0.00 & [-0.04, 0.05] & 58.23\% & 73.80\% & 25,636 \\
\addlinespace[8pt]
\textbf{Visual Inter. × Impulsivity} & \textbf{-0.29} & \textbf{[-0.56, -0.03]} & \textbf{98.69\%} & \textbf{0\%} & \textbf{19,785} \\
Haptic Inter. × Impulsivity & -0.17 & [-0.41, 0.06] & 93.03\% & 5.79\% & 23,975 \\
\addlinespace[4pt]
Visual Inter. × FOMO & 0.00 & [-0.12, 0.13] & 53.83\% & 31.42\% & 22,350 \\
Haptic Inter. × FOMO & 0.06 & [-0.05, 0.17] & 84.44\% & 21.43\% & 27,067 \\
\addlinespace[4pt]
Visual Inter. × Self-Control & 0.08 & [-0.15, 0.30] & 75.25\% & 14.23\% & 18,754 \\
Haptic Inter. × Self-Control & 0.05 & [-0.15, 0.26] & 69.58\% & 17.31\% & 21,192 \\
\addlinespace[4pt]
\textbf{Visual Inter. × Anxiety} & \textbf{0.12} & \textbf{[0.00, 0.23]} & \textbf{97.81\%} & \textbf{2.82\%} & \textbf{24,627} \\
Haptic Inter. × Anxiety & 0.00 & [-0.11, 0.11] & 50.89\% & 35.80\% & 30,901 \\
\midrule
\addlinespace[2pt]
\multicolumn{6}{l}{$Conditional~R^2 = 0.80$ (95\% CrI $[0.79, 0.81]$)} \\
\multicolumn{6}{l}{$Marginal~R^2 = 0.28$, (95\% CrI $[0.21, 0.35]$} \\
\bottomrule

\end{tabular}
\end{table*}

\begin{table*}[ht]
\small
\caption{Results of the Bayesian mixed-effects model for \textbf{objective effectiveness}. Effects with a \textit{pd} $>$ 95\% are highlighted in bold.}
\label{tab:results_objective}
\begin{tabular}{lccccc}
\toprule
\toprule
\multicolumn{6}{c}{\textbf{Objective Effectiveness}} \\
\toprule
\toprule
\addlinespace[4pt]
Parameter & Median & 95\% CrI & pd & \% in ROPE & ESS \\
\midrule
\textit{Intercept} & \textit{0.83} & \textit{[0.48, 1.19]} & \textit{100\%} & \textit{0.00\%} & \textit{11,143} \\
\addlinespace[4pt]
\textbf{Visual Inter}. &\textbf{ 0.36} & \textbf{[0.01, 0.72]} & \textbf{97.67\%} & \textbf{0.83\%} & \textbf{17,219} \\
Haptic Inter. & 0.03 & [-0.31, 0.37] & 55.77\% & 13.04\% & 18,762 \\
\addlinespace[4pt]
Valence & -0.07 & [-0.21, 0.06] & 86.42\% & 18.99\% & 15,431 \\
Sleepiness & -0.08 & [-0.18, 0.02] & 94.89\% & 12.61\% & 19,957 \\
Stress & -0.02 & [-0.16, 0.12] & 61.79\% & 30.37\% & 17,187 \\
Multitasking [Yes] & 0.03 & [-0.03, 0.09] & 80.52\% & 50.68\% & 21,658 \\
At Home [True] & 0.01 & [-0.08, 0.09] & 55.09\% & 49.31\% & 12,706 \\
Current Activity & 0.04 & [-0.06, 0.13] & 79.05\% & 34.84\% & 17,694 \\
Social Situation [Friends] & -0.04 & [-0.11, 0.02] & 90.70\% & 31.62\% & 21,886 \\
\addlinespace[4pt]
Impulsivity & 0.06 & [-0.34, 0.46] & 61.86\% & 11.21\% & 12,324 \\
FOMO & 0.00 & [-0.21, 0.20] & 51.79\% & 22.57\% & 13,987 \\
Self-Control & -0.19 & [-0.52, 0.16] & 85.20\% & 7.49\% & 12,573 \\
Anxiety & -0.06 & [-0.25, 0.13] & 73.47\% & 20.13\% & 17,986 \\
\addlinespace[2pt]
\hline
\addlinespace[2pt]
Visual Inter. × Valence & 0.07 & [-0.11, 0.24] & 78.06\% & 18.93\% & 17,218 \\
Haptic Inter. × Valence & 0.12 & [-0.05, 0.29] & 90.88\% & 11.15\% & 14,751 \\
\addlinespace[4pt]
\textbf{Visual Inter. × Sleepiness} & \textbf{0.13} & \textbf{[0.01, 0.26]} & \textbf{98.02\%} & \textbf{2.83\%} & \textbf{19,762} \\
Haptic Inter. × Sleepiness & 0.08 & [-0.04, 0.21] & 89.93\% & 16.19\% & 21,316 \\
\addlinespace[4pt]
Visual Inter. × Stress & 0.07 & [-0.12, 0.25] & 75.94\% & 19.47\% & 17,595 \\
Haptic Inter. × Stress & 0.00 & [-0.19, 0.17] & 53.60\% & 25.19\% & 16,668 \\
\addlinespace[4pt]
Visual Inter. × Multitasking [Yes] & 0.02 & [-0.06, 0.10] & 70.20\% & 48.98\% & 21,934 \\
Haptic Inter. × Multitasking [Yes] & 0.03 & [-0.05, 0.11] & 77.85\% & 41.59\% & 21,536 \\
\addlinespace[4pt]
Visual Inter. × At Home [True] & 0.03 & [-0.09, 0.14] & 66.31\% & 34.76\% & 18,697 \\
Haptic Inter. × At Home [True] & 0.05 & [-0.07, 0.17] & 81.37\% & 25.68\% & 14,291 \\
\addlinespace[4pt]
Visual Inter. × FOMO & 0.02 & [-0.22, 0.25] & 54.97\% & 19.02\% & 14,414 \\
Haptic Inter. × FOMO & 0.12 & [-0.10, 0.35] & 85.88\% & 11.28\% & 15,072 \\
\addlinespace[4pt]
Visual Inter. × Current Activity & -0.04 & [-0.16, 0.09] & 72.55\% & 30.20\% & 18,940 \\
Haptic Inter. × Current Activity & 0.00 & [-0.12, 0.13] & 52.12\% & 35.02\% & 20,771 \\
\addlinespace[4pt]
Visual Inter. × Social Situation [Friends] & 0.02 & [-0.06, 0.10] & 69.94\% & 46.03\% & 23399 \\
Haptic Inter. × Social Situation [Friends] & 0.02 & [-0.06, 0.10] & 67.55\% & 46.07\% & 23216 \\
\addlinespace[8pt]
Visual inter. × Impulsivity & -0.04 & [-0.52, 0.43] & 56.00\% & 9.82\% & 13,301 \\
\textbf{Haptic Inter. × Impulsivity} & \textbf{0.42} & \textbf{[-0.03, 0.86]} & \textbf{96.61\%} & \textbf{1.91\%} & \textbf{13,337} \\
\addlinespace[4pt]
Visual Inter. × Self-Control & 0.30 & [-0.11, 0.71] & 92.58\% & 3.92\% & 13,539 \\
\textbf{Haptic Inter. × Self-Control} & \textbf{0.58} & \textbf{[0.18, 0.97]} & \textbf{99.74\%} & \textbf{0.00\%} & \textbf{11,703} \\
\addlinespace[4pt]
Visual Inter. × Anxiety & 0.03 & [-0.19, 0.25] & 60.27\% & 20.16\% & 17,304 \\
Haptic inter. × Anxiety & 0.02 & [-0.19, 0.24] & 58.59\% & 20.26\% & 18,816 \\
\midrule
\addlinespace[2pt]
\multicolumn{6}{l}{$Conditional~R^2 = 0.45$ (95\% CrI $[0.40, 0.50]$)} \\
\multicolumn{6}{l}{$Marginal~R^2 = 0.12$, (95\% CrI $[0.09, 0.17]$)} \\
\bottomrule

\end{tabular}
\end{table*}

\clearpage
 
\section{Sensitivity Analysis}\label{app:sensitivity}
 
To assess the robustness of the moderation effects reported in the main analysis, we re-calculated both Bayesian mixed-effects models under two alternative participant filtering conditions. The primary analysis retained only participants who experienced all three intervention types ($N=104$). In the first sensitivity condition, we relaxed this criterion to include all participants who completed at least three interventions in total, regardless of whether all three types were represented ($N=150$). In the second condition, we included all 214 participants who completed the 7-day study, with no filtering based on intervention exposure. In both sensitivity conditions, the same preprocessing pipeline (outlier removal, normalization, model specification, and priors) was applied. \autoref{tab:sensitivity} shows the results for all interaction effects that were possibly existing ($pd > 95\%$) in at least one of the three filtering conditions. 
The results show that the anxiety and self-control interaction effects were robust across all three conditions, with significance strengthening as more participants were included. The impulsivity interactions attenuated but remained directionally consistent, which is expected given the reduced within-person balance in the larger samples. Several additional effects emerged only under relaxed filtering (e.g., \textit{Visual Inter. × Self-Control}, \textit{Haptic Inter. × FoMo}) and should be interpreted cautiously. Overall, these results support the primary filtering decision while demonstrating that the core findings are not artifacts of this particular filtering threshold.
 
\begin{table*}[ht]
\setlength{\tabcolsep}{5pt}
\renewcommand{\arraystretch}{0.92}
\small
\caption{Sensitivity analysis across three participant filtering conditions. Effects with $pd > 95\%$ are bold.}
\label{tab:sensitivity}
\begin{tabular}{llcccc}
\midrule
\midrule
\multicolumn{6}{c}{\textbf{Subjective Effectiveness}} \\
\midrule
\midrule
Interaction Effect & Filtering & Median & 95\% CrI & pd & \% in ROPE \\
\midrule
\multirow{3}{*}{Visual Inter. $\times$ Impulsivity}
  & $N=104$ (primary)       & \textbf{$-$0.29} & \textbf{[$-$0.56, $-$0.03]} & \textbf{98.69\%} & \textbf{0.00\%} \\
  & $N=150$ (intermediate)  & $-$0.11 & [$-$0.33, 0.12] & 82.91\% & 10.66\% \\
  & $N=214$ (relaxed)       & $-$0.07 & [$-$0.28, 0.14] & 74.52\% & 14.67\% \\
\addlinespace[3pt]
\multirow{3}{*}{Visual Inter. $\times$ Anxiety}
  & $N=104$ (primary)       & \textbf{0.12} & \textbf{[0.00, 0.23]} & \textbf{97.81\%} & \textbf{2.82\%} \\
  & $N=150$ (intermediate)  & \textbf{0.15} & \textbf{[0.04, 0.26]} & \textbf{99.61\%} & \textbf{0.00\%} \\
  & $N=214$ (relaxed)       & \textbf{0.15} & \textbf{[0.05, 0.25]} & \textbf{99.74\%} & \textbf{0.00\%} \\
\addlinespace[3pt]
\multirow{3}{*}{Visual Inter. $\times$ Self-Control}
  & $N=104$ (primary)       & 0.08 & [$-$0.15, 0.30] & 75.25\% & 14.23\% \\
  & $N=150$ (intermediate)  & 0.16 & [$-$0.04, 0.36] & 93.73\% & 6.08\% \\
  & $N=214$ (relaxed)       & \textbf{0.17} & \textbf{[$-$0.02, 0.36]} & \textbf{96.33\%} & \textbf{3.69\%} \\
\addlinespace[1pt]
\midrule
\midrule
\multicolumn{6}{c}{\textbf{Objective Effectiveness}} \\
\midrule
\midrule
\multirow{3}{*}{Haptic Inter. $\times$ Self-Control}
  & $N=104$ (primary)       & \textbf{0.58} & \textbf{[0.18, 0.97]} & \textbf{99.74\%} & \textbf{0.00\%} \\
  & $N=150$ (intermediate)  & \textbf{0.61} & \textbf{[0.26, 0.96]} & \textbf{99.98\%} & \textbf{0.00\%} \\
  & $N=214$ (relaxed)       & \textbf{0.60} & \textbf{[0.26, 0.93]} & \textbf{99.97\%} & \textbf{0.00\%} \\
\addlinespace[3pt]
\multirow{3}{*}{Haptic Inter. $\times$ Impulsivity}
  & $N=104$ (primary)       & \textbf{0.42} & \textbf{[$-$0.03, 0.86]} & \textbf{96.61\%} & \textbf{1.91\%} \\
  & $N=150$ (intermediate)  & 0.30 & [$-$0.09, 0.69] & 93.77\% & 3.49\% \\
  & $N=214$ (relaxed)       & 0.28 & [$-$0.10, 0.65] & 92.88\% & 3.98\% \\
\addlinespace[3pt]
\multirow{3}{*}{Visual Inter. $\times$ Sleepiness}
  & $N=104$ (primary)       & \textbf{0.13} & \textbf{[0.01, 0.26]} & \textbf{98.02\%} & \textbf{2.83\%} \\
  & $N=150$ (intermediate)  & \textbf{0.11} & \textbf{[$-$0.01, 0.23]} & \textbf{96.72\%} & \textbf{5.96\%} \\
  & $N=214$ (relaxed)       & 0.09 & [$-$0.04, 0.21] & 91.65\% & 14.68\% \\
\addlinespace[3pt]
\multirow{3}{*}{Visual Inter. $\times$ Self-Control}
  & $N=104$ (primary)       & 0.30 & [$-$0.11, 0.71] & 92.58\% & 3.92\% \\
  & $N=150$ (intermediate)  & \textbf{0.33} & \textbf{[0.00, 0.67]} & \textbf{96.94\%} & \textbf{1.92\%} \\
  & $N=214$ (relaxed)       & \textbf{0.34} & \textbf{[0.00, 0.67]} & \textbf{97.49\%} & \textbf{1.23\%} \\
\addlinespace[3pt]
\multirow{3}{*}{Haptic Inter. $\times$ FoMo}
  & $N=104$ (primary)       & 0.12 & [$-$0.10, 0.35] & 85.88\% & 11.28\% \\
  & $N=150$ (intermediate)  & \textbf{0.20} & \textbf{[0.00, 0.40]} & \textbf{97.32\%} & \textbf{2.41\%} \\
  & $N=214$ (relaxed)       & \textbf{0.23} & \textbf{[0.04, 0.43]} & \textbf{99.04\%} & \textbf{0.00\%} \\
\bottomrule
\end{tabular}
\end{table*}

%TC:endignore

\end{document}